\documentstyle[12pt]{article}
\begin{document}


\def\Nequalstwo{\Psi}
\def\eff{{\rm eff}}
\def\inst{{\rm inst}}
\def\fermi{{\rm fermi}}
\def\trtwo{\tr^{}_2\,}
\def\finv{f^{-1}}
\def\Ubar{\bar U}
\def\wbar{\bar w}
\def\fbar{\bar f}
\def\abar{\bar a}
\def\bbar{\bar b}
\def\Deltabar{\bar\Delta}
\def\dalpha{{\dot\alpha}}
\def\dbeta{{\dot\beta}}
\def\dgamma{{\dot\gamma}}
\def\ddelta{{\dot\delta}}
\def\Sbar{\bar S}
\def\Im{{\rm Im}}
\def\sst{\scriptscriptstyle}
\def\cld{C_{\sst\rm LD}^{}}
\def\csd{C_{\sst\rm SD}^{}}
\def\bigI{{\rm I}_{\sst 3\rm D}}
\def\Mr{{\rm M}_{\sst R}}
\def\cJ{C_{\sst J}}
\def\one{{\sst(1)}}
\def\two{{\sst(2)}}
\def\vsd{v^{\sst\rm SD}}
\def\vasd{v^{\sst\rm ASD}}
\def\Phibar{\bar\Phi}
\def\F{{\cal F}_{\sst\rm SW}}
\def\P{{\cal P}}
\def\A{{\cal A}}
\def\susy{supersymmetry}
\def\sigmabar{\bar\sigma}
\def\barsigma{\sigmabar}
\def\ASD{{\scriptscriptstyle\rm ASD}}
\def\cl{{\,\rm cl}}
\def\lambdabar{\bar\lambda}
\def\R{{R}}
\def\psibar{\bar\psi}
\def\sqrtwo{\sqrt{2}\,}
\def\etabar{\bar\eta}
\def\Thetabar{{\bar\Theta_0}}
\def\Qbar{\bar Q}
\def\susic{supersymmetric}
\def\vhiggs{{\rm v}}
\def\vhiggsa{{\cal A}_{\sst00}}
\def\vbarhiggs{\bar{\rm v}}
\def\vhiggsbar{\bar{\rm v}}
\def\novetal{Novikov et al.}
\def\Novetal{Novikov et al.}
\def\ADS{Affleck, Dine and Seiberg}
\def\ads{Affleck, Dine and Seiberg}
\def\setI{\{{\cal I}\}}
\def\Abar{A^\dagger}
\def\B{{\cal B}}
\def\infinity{\infty}
\def\C{{\cal C}}
\def\Psitwo{\Psi_{\scriptscriptstyle N=2}}
\def\Psibartwo{\bar\Psi_{\scriptscriptstyle N=2}}
\def\zero{{\scriptscriptstyle(0)}}
\def\new{{\scriptscriptstyle\rm new}}
\def\u{\underline}
\def\uA{\,\lower 1.2ex\hbox{$\sim$}\mkern-13.5mu A}
\def\uBmu{\,\lower 1.2ex\hbox{$\sim$}\mkern-13.5mu B_\mu}
\def\uAmu{\,\lower 1.2ex\hbox{$\sim$}\mkern-13.5mu A_\mu}
\def\uX{\,\lower 1.2ex\hbox{$\sim$}\mkern-13.5mu X}
\def\uD{\,\lower 1.2ex\hbox{$\sim$}\mkern-13.5mu {\rm D}}
\def\uDzero{{\uD}^\zero}
\def\uAzero{{\uA}^\zero}
\def\upsizero{{\upsi}^\zero}
\def\uF{\,\lower 1.2ex\hbox{$\sim$}\mkern-13.5mu F}
\def\uW{\,\lower 1.2ex\hbox{$\sim$}\mkern-13.5mu W}
\def\uWbar{\,\lower 1.2ex\hbox{$\sim$}\mkern-13.5mu {\overline W}}
\def\Dbar{D^\dagger}
\def\Fbar{F^\dagger}
\def\uAbar{{\uA}^\dagger}
\def\uAbarzero{{\uA}^{\dagger\zero}}
\def\uDbar{{\uD}^\dagger}
\def\uDbarzero{{\uD}^{\dagger\zero}}
\def\uFbar{{\uF}^\dagger}
\def\uFbarzero{{\uF}^{\dagger\zero}}
\def\uV{\,\lower 1.2ex\hbox{$\sim$}\mkern-13.5mu V}
\def\uZ{\,\lower 1.2ex\hbox{$\sim$}\mkern-13.5mu Z}
\def\uv{\lower 1.0ex\hbox{$\scriptstyle\sim$}\mkern-11.0mu v}
\def\uc{\lower 1.0ex\hbox{$\scriptstyle\sim$}\mkern-11.0mu c}
\def\uB{\lower 1.0ex\hbox{$\scriptstyle\sim$}\mkern-11.0mu B}
\def\uPsi{\,\lower 1.2ex\hbox{$\sim$}\mkern-13.5mu \Psi}
\def\uPhi{\,\lower 1.2ex\hbox{$\sim$}\mkern-13.5mu \Phi}
\def\uchi{\lower 1.5ex\hbox{$\sim$}\mkern-13.5mu \chi}
\def\utheta{\lower 1.5ex\hbox{$\sim$}\mkern-13.5mu \theta}
\def\chitilde{\tilde \chi}
\def\etatilde{\tilde \eta}
\def\uchitilde{\lower 1.5ex\hbox{$\sim$}\mkern-13.5mu \tilde\chi}
\def\ueta{\lower 1.5ex\hbox{$\sim$}\mkern-13.5mu \eta}
\def\uetatilde{\lower 1.5ex\hbox{$\sim$}\mkern-13.5mu \tilde\eta}
\def\Psibar{\bar\Psi}
\def\uPsibar{\,\lower 1.2ex\hbox{$\sim$}\mkern-13.5mu \Psibar}
\def\upsi{\,\lower 1.5ex\hbox{$\sim$}\mkern-13.5mu \psi}
\def\uphi{\lower 1.5ex\hbox{$\sim$}\mkern-13.5mu \phi}
\def\uphione{\lower 1.5ex\hbox{$\sim$}\mkern-13.5mu \phi_1}
\def\uph2{\lower 1.5ex\hbox{$\sim$}\mkern-13.5mu \phi_2}
\def\psibar{\bar\psi}
\def\upsibar{\,\lower 1.5ex\hbox{$\sim$}\mkern-13.5mu \psibar}
\def\etabar{\bar\eta}
\def\uetabar{\,\lower 1.5ex\hbox{$\sim$}\mkern-13.5mu \etabar}
\def\chibar{\bar\chi}
\def\uchibar{\,\lower 1.5ex\hbox{$\sim$}\mkern-13.5mu \chibar}
\def\upsibarzero{\,\lower 1.5ex\hbox{$\sim$}\mkern-13.5mu \psibar^\zero}
\def\ulambda{\,\lower 1.2ex\hbox{$\sim$}\mkern-13.5mu \lambda}
\def\ulambdabar{\,\lower 1.2ex\hbox{$\sim$}\mkern-13.5mu \lambdabar}
\def\ulambdabarzero{\,\lower 1.2ex\hbox{$\sim$}\mkern-13.5mu \lambdabar^\zero}
\def\ulambdabarnew{\,\lower 1.2ex\hbox{$\sim$}\mkern-13.5mu \lambdabar^\new}
\def\D{{\cal D}}
\def\M{{\cal M}}
\def\N{{\cal N}}
\def\Dslash{\,\,{\raise.15ex\hbox{/}\mkern-12mu D}}
\def\Dbarslash{\,\,{\raise.15ex\hbox{/}\mkern-12mu {\bar D}}}
\def\delslash{\,\,{\raise.15ex\hbox{/}\mkern-9mu \partial}}
\def\delbarslash{\,\,{\raise.15ex\hbox{/}\mkern-9mu {\bar\partial}}}
\def\L{{\cal L}}
\def\hf{{\textstyle{1\over2}}}
\def\quarter{{\textstyle{1\over4}}}
\def\twe{{\textstyle{1\over12}}}
\def\eighth{{\textstyle{1\over8}}}
\def\fourth{\quarter}
\def\wb{Wess and Bagger}
\def\xibar{\bar\xi}
\def\ss{{\scriptscriptstyle\rm ss}}
\def\sc{{\scriptscriptstyle\rm sc}}
\def\uvcl{{\uv}^\cl}
\def\uAcl{\,\lower 1.2ex\hbox{$\sim$}\mkern-13.5mu A^{}_{\cl}}
\def\uAbarcl{\,\lower 1.2ex\hbox{$\sim$}\mkern-13.5mu A_{\cl}^\dagger}
\def\upsinew{{\upsi}^\new}
\def\ASDzero{{{\scriptscriptstyle\rm ASD}\zero}}
\def\SDzero{{{\scriptscriptstyle\rm SD}\zero}}
\def\SD{{\scriptscriptstyle\rm SD}}
\def\varthetabar{{\bar\vartheta}}
\def\three{{\scriptscriptstyle(3)}}
\def\dagthree{{\dagger\scriptscriptstyle(3)}}
\def\ld{{\scriptscriptstyle\rm LD}}
\def\vld{v^\ld}
\def\Dld{{\rm D}^\ld}
\def\Fld{F^\ld}
\def\Ald{A^\ld}
\def\Fbarld{F^{\dagger\scriptscriptstyle\rm LD}}
\def\Abarld{A^{\dagger\scriptscriptstyle \rm LD}}
\def\lambdald{\lambda^\ld}
\def\lambdabarld{\bar\lambda^\ld}
\def\psild{\psi^\ld}
\def\psibarld{\bar\psi^\ld}
\def\dsiginst{d\sigma_{\scriptscriptstyle\rm inst}}
\def\xione{\xi_1}
\def\xionebar{\bar\xi_1}
\def\xitwo{\xi_2}
\def\xitwobar{\bar\xi_2}
\def\thetatwo{\vartheta_2}
\def\thetatwobar{\bar\vartheta_2}
\def\Ltwo{\L_{\sst SU(2)}}
\def\Leff{\L_{\rm eff}}
\def\Laux{\L_{\rm aux}}
\def\oneloop{{\sst\rm 1\hbox{-}\sst\rm loop}}
\def\LSUtwo{{\cal L}_{\rm SU(2)}}
\def\Dhat{\hat\D}
\def\bkgd{{\sst\rm bkgd}}
\def\Lgft{{\cal L}_{\sst\rm g.f.t.}}
\def\Lghost{{\cal L}_{\sst\rm ghost}}
\def\Sinst{S_{\rm inst}}
\def\etal{{\rm et al.}}
\def\S{{\cal S}}
\def\L{ {\cal L}}
\def\C{ {\cal C}}
\def\N{ {\cal N}}
\def\calE{{\cal E}}
\def\lin{{\rm lin}}
\def\Tr{{\rm Tr}}
\def\mxth{\mathsurround=0pt }
\def\xversim#1#2{\lower2.pt\vbox{\baselineskip0pt \lineskip-.5pt
x  \ialign{$\mxth#1\hfil##\hfil$\crcr#2\crcr\sim\crcr}}}
\def\simgr{\mathrel{\mathpalette\xversim >}}
\def\simle{\mathrel{\mathpalette\xversim <}}
\def\slash{\llap /}
\def\lagr{{\cal L}}

\renewcommand{\a}{\alpha}
\renewcommand{\b}{\beta}
\renewcommand{\c}{\gamma}
\renewcommand{\d}{\delta}
\newcommand{\pa}{\partial}
\newcommand{\g}{\gamma}
\newcommand{\G}{\Gamma}
\newcommand{\e}{\epsilon}
\newcommand{\z}{\zeta}
\newcommand{\Z}{\Zeta}
\newcommand{\K}{\Kappa}
\renewcommand{\l}{\lambda}
\renewcommand{\L}{\Lambda}
\newcommand{\m}{\mu}
\newcommand{\n}{\nu}
\newcommand{\X}{\Chi}

\newcommand{\s}{\sigma}
\renewcommand{\S}{\Sigma}
\renewcommand{\t}{\tau}
\newcommand{\T}{\Tau}
\newcommand{\y}{\upsilon}
\newcommand{\Y}{\upsilon}
\renewcommand{\o}{\omega}
\newcommand{\q}{\theta}
\newcommand{\h}{\eta}
\newcommand{\cmap}{{$\bf c$} map}
\newcommand{\Ka}{K\"ahler} 
\renewcommand{\O}{{\Omega}}
\newcommand{\var}{\varepsilon}
%

\newcommand{\nd}[1]{/\hspace{-0.5em} #1}
\begin{titlepage}
\begin{flushright}
{\bf June 1999} \\ 
UW/PT 99-10 \\ 
hep-th/9906011 \\
\end{flushright}
\begin{centering}
\vspace{.2in}
{\large {\bf  An Elliptic Superpotential for Softly Broken  \\ ${\cal N}=4$ 
Supersymmetric Yang-Mills Theory }}\\
\vspace{.4in}
 N. Dorey \\
\vspace{.4in}
Department of Physics, University of Washington, Box 351560   \\
Seattle, Washington 98195-1560, USA\\
\vspace{.2in}
and \\ 
\vspace{.2in}
Department of Physics, University of Wales Swansea \\
Singleton Park, Swansea, SA2 8PP, UK\\
\vspace{.4in}
{\bf Abstract} \\
\end{centering}
An exact superpotential is derived for the ${\cal N}=1$ theories which 
arise as massive deformations of ${\cal N}=4$ supersymmetric Yang-Mills 
(SYM) theory. The superpotential of the $SU(N)$ theory formulated on 
$R^{3}\times S^{1}$ is shown to coincide with the complexified 
potential of the $N$-body elliptic Calogero-Moser Hamiltonian. 
This superpotential reproduces the vacuum structure predicted by 
Donagi and Witten for the corresponding four-dimensional theory and 
also transforms covariantly under the S-duality group of ${\cal N}=4$ 
SYM. The analysis yields exact formulae with interesting modular 
properties for the condensates of gauge-invariant chiral operators 
in the four-dimensional theory. 


\end{titlepage}

\section{Introduction and Review}
\paragraph{} 

Four-dimensional gauge theories with ${\cal N}=4$ 
supersymmetry are believed to provide an exact realization
of the duality between electric and magnetic charges 
originally proposed by Montonen and Olive \cite{OM}. The extent 
to which this duality is preserved in theories with less supersymmetry, 
obtained as massive deformations of the 
${\cal N}=4$ theory, is an important and interesting question. 
For deformations which preserve ${\cal N}=2$ supersymmetry, 
the exact solution of the corresponding low-energy theory is determined by a 
complex curve which is manifestly invariant under $SL(2,Z)$ 
transformations. The relevant curve was 
given by Seiberg and Witten for gauge group $SU(2)$ 
\cite{SW2}, and subsequently generalized to the $SU(N)$ case 
by Donagi and Witten \cite{DW}. The singularities of these 
curves also determine 
the vacuum structure of the theories obtained by breaking 
${\cal N}=4$ supersymmetry down to an ${\cal N}=1$ subalgebra. 
In terms of ${\cal N}=1$ supersymmetry, these theories contain a 
vector multiplet and three massive adjoint chiral multiplets.  
Donagi and Witten used this correspondence to show that 
the ${\cal N}=1$ theory with gauge group $SU(N)$ has a total of 
$\sum_{d|N}d$ massive vacua which are permuted by S-duality 
transformations. For $N>2$, the theory also has massless vacua. 
\paragraph{}
In this paper, I will present an alternative analysis of the ${\cal N}=1$ 
theories described above. 
The approach followed, based on that of \cite{SW3}, is to compactify 
the four-dimensional theory on a circle of radius $R$. As we review below, 
the $SU(N)$ theory then has a Coulomb branch parametrized by 
$N$ chiral superfields\footnote{Some conventions used throughout this 
paper: Cartan algebra valued fields for $SU(N)$ such as $X=X^{a}H^{a}$ have  
$N$ components $X^{a}$, $a=1,\ldots, N$, which are subject to the constraint 
$\sum_{a=1}^{N}X^{a}=0$. The only exception is for gauge group $SU(2)$ where 
we work with a single unconstrained field $X=X^{2}-X^{1}$. 
Supersymmetries 
are counted with four-dimensional conventions throughout. Thus 
${\cal N}=1$ always denotes a theory with four supercharges.}, 
$X^{a}$, which take values on a torus whose 
complex structure parameter $\tau$ is the complexified gauge coupling. 
The main result is that the exact superpotential of the theory is, 
\begin{equation} 
{\cal W}= m_{1}m_{2}m_{3}\, \sum_{a>b} \, {\cal P}(X_{a}-X_{b})
\label{sp1}
\end{equation} 
where $m_{i}$, $i=1,2,3$ are the masses of the adjoint chiral multiplets 
and ${\cal P}$ is the Weierstrass function \cite{EF}.  
Note that ${\cal W}$ 
does not depend on $R$. In the limit $R\rightarrow \infty$, 
comparison with the four-dimensional results of \cite{DW} provides 
a detailed test of this proposal. 
In particular, we will see that ${\cal W}$ has a set of critical points 
which match the singular points of the Donagi-Witten curves. The 
superpotential (\ref{sp1}) transforms with weight two 
under $SL(2,Z)$ modular transformations and the action of S-duality 
on the massive vacua predicted in \cite{DW} is manifest in this approach.  
\paragraph{}
For massive theories with ${\cal N}=1$ supersymmetry we typically expect 
to obtain exact results for the condensates of gauge-invariant chiral 
operators \cite{AK,NSVZ}. 
Where a description in terms of soft breaking of ${\cal N}=2$ 
supersymmetry exists, such results can be obtained directly from the 
corresponding complex curve. For example consider the 
condensates $u_{n}=\langle {\rm Tr}\Phi^{n}\rangle$, $n=2,\ldots,N$, 
where $\Phi$ is one of the adjoint scalar fields of the $SU(N)$ theory. 
In principle, the values of the $u_{n}$ in 
each vacuum be infered from the location 
of the corresponding singular point 
in the moduli space of the Donagi-Witten curve. 
In practice, this is approach is not straightforward because 
there is no explicit formula for the $SU(N)$ curve.     
As a simple application of the superpotential described above, 
I calculate the exact value of the condensate $u_{2}$ in each massive 
vacuum of the $SU(N)$ theory when $N$ is prime. 
The resulting set of vacuum values transforms with weight 
two under $SL(2,Z)$, modulo the expected permutations of the vacua. 
For a related approach to calculating 
the gluino condensate in ${\cal N}=1$ supersymmetric 
Yang-Mills theory see \cite{HKM}. Additional recent discussion of the 
superpotential of ${\cal N}=1$ SYM in four dimensions 
is given in \cite{GA}. Another interesting aspect of these results is 
that the superpotential (\ref{sp1}) coincides 
with the (complexified) potential of the elliptic Calogero-Moser Hamiltonian. 
The latter is the integrable system associated with the Donagi-Witten 
solution of the four-dimensional ${\cal N}=2$ theory mentioned above 
\cite{MA}. This reveals an interesting connection between ${\cal N}=1$ 
theories and integrability which is sketched in the final section of 
the paper.       
\subsection{The four-dimensional theory}
\paragraph{}
We begin by considering ${\cal N}=4$ supersymmetric Yang-Mills (SYM) 
theory with gauge group $SU(N)$ in four dimensions. In terms of ${\cal N}=1$ 
superfields this theory contains a gauge multiplet $V$ as 
well as three chiral multiplets $\Phi_{i}$, $i=1,2,3$ in the adjoint 
representation of $SU(N)$. 
The superspace Lagrangian includes a superpotential 
$W={\rm Tr}(\Phi_{1}[\Phi_{2},\Phi_{3}])$ which preserves an $SU(4)$ 
R-symmetry but explicitly breaks the potentially anomalous global $U(1)$ 
under which the $\Phi_{i}$ each transform with charge $+1$. 
Soft breaking to ${\cal N}=1$ is accomplished
by introducing masses $m_{i}$ for each chiral superfield $\Phi_{i}$. 
Including these terms the superpotential becomes, 
\begin{equation}
W={\rm Tr}\left(\Phi_{1}[\Phi_{2},\Phi_{3}]+ 
m_{1}\Phi_{1}^{2}+ m_{2}\Phi_{2}^{2}+m_{3}\Phi_{3}^{2}\right)
\label{spmass}
\end{equation}
For generic values of the masses $m_{i}$, the $SU(4)$ R-symmetry is 
completely broken. It is convenient to rescale the chiral superfields 
as $\Phi_{1}=2\sqrt{m_{2}m_{3}} X$,  $\Phi_{2}=2\sqrt{m_{3}m_{1}} Y$, 
$\Phi_{3}=-2\sqrt{m_{1}m_{2}} Z$. Up to an overall factor, the 
superpotential becomes, 
\begin{equation}
W={\rm Tr}\left(\frac{1}{2}(X^{2}+Y^{2}+Z^{2})-X[Y,Z] \right)
\label{spxyz}
\end{equation}
\paragraph{}
Following \cite{VW,DW}, we will now determine vacuum structure of 
the classical theory with superpotential (\ref{spmass}). To do this 
we must impose the F-term equations 
which come from varying $W$ with respect to $X$, $Y$ and $Z$. 
The first equation is $X=[Y,Z]$ and the other two are generated by 
cyclic permutation of the three superfields. A supersymmetric vacuum 
is therefore described by three $N\times N$ matrices which obey the 
commutation relations of an $SU(2)$ Lie algebra. However, we still 
have to impose the corresponding D-term equations and mod out by $SU(N)$ gauge 
transformations. As usual these two steps can be combined by dividing out 
the action of the complexified gauge group $SL(N,C)$ on $X$, $Y$ and $Z$. 
In fact, up to an $SL(N,C)$ gauge transformation, there is exactly one 
solution of the $SU(2)$ commutation relations in terms of $N\times N$ matrices 
for each $N$-dimensional representation of $SU(2)$. Except for  
the unique irreducible representation of dimension $N$, each 
representation can be decomposed as the sum of irreducible pieces 
of dimensions $n_{1},n_{2},\ldots n_{r}$ with $\sum_{l}n_{l}=N$.  
In general, unless 
the representation is trivial, the corresponding non-zero values of 
the adjoint scalar fields break the gauge group $SU(N)$ down to a subgroup 
$H$. The vacuum structure for each representation  $\rho$
is essentially determined by $H$ and may be analysed as follows:     
\paragraph{}
{\bf 1} If $\rho$ is the irreducible representation, 
the gauge group is completely broken. In this case the 
theory is in a Higgs phase with a mass gap. As the theory remains weakly 
coupled at all energy scales, the semiclassical analysis can be taken at 
face value and we find exactly one vacuum.     
\paragraph{}
{\bf 2} If $\rho$ is trivial, the gauge group is completely unbroken 
and the theory is in a confining phase.  
At low energies, the adjoint matter multiplets decouple and the 
theory flows to ${\cal N}=1$ SYM. The effective theory therefore becomes  
strongly-coupled in the IR and generates mass gap. In particular 
${\cal N}=1$ SYM 
has an $Z_{2N}$ global symmetry 
which is spontaneously broken to $Z_{2}$ by gluino condensation giving  
$N$ supersymmetric vacua \cite{AK,NSVZ}. 
Cases where $\rho$ is the sum of $d$ irreducible 
representations each of the same dimension also lead to confinement. In 
these cases $H=SU(d)$ and the theory flows to ${\cal N}=1$ SYM with 
this gauge group. By similar reasoning this gives $d$ massive 
supersymmetric vacua. 
\paragraph{}
{\bf 3} In the remaining cases $\rho$ includes irreducible pieces of different 
dimensions and the unbroken gauge group includes abelian factors. This 
means that the theory is realized in a Coulomb phase with at least one 
massless photon.       
\paragraph{}
Combining case {\bf 1} and {\bf 2} we find that the total number of massive 
vacua of the $SU(N)$ theory is the sum of the divisors of $N$. The case 
where $N$ is prime is particularly straightforward as only the irreducible 
and the trivial representations contribute and hence there are exactly 
$N+1$ massive vacua. In particular, the above analysis reveals that the 
$SU(2)$ theory has three massive vacua, one in the Higgs phase and two in 
the confinement phase. The possibility of massless vacua first arises 
for gauge group 
$SU(3)$ where $\rho$ can be chosen as a sum of one copy of 
the trivial representation and one of the fundamental. In this configuration 
the unbroken gauge group is $U(1)\times SU(2)$. Donagi and Witten 
showed that choice 
leads to a single supersymmetric vacuum in the Coulomb phase.        

\subsection{Compactification to Three Dimensions}
\paragraph{}
We now consider what happens if one compactifies the four-dimensional theory 
introduced above to three dimensions on a circle of radius $R$. The 
most obvious modification is that the theory 
acquires new scalar degrees of freedom which come from the Wilson 
lines of the four-dimensional gauge field around the compact dimension. 
Non-zero values for these scalars generically break the gauge group 
down to its maximal abelian subalgebra; $SU(N)\rightarrow U(1)^{N-1}$. 
Hence, at the classical level, the compactified theory has a Coulomb branch. 
By a gauge rotation, the Wilson line can be chosen to 
lie in the Cartan subalgebra, $\phi=\phi^{a}H^{a}$. 
Each scalar $\phi^{a}$ is related to the corresponding 
Cartan component of the four-dimensional gauge field   
as $\phi^{a}=\int_{S^{1}} A^{a}\cdot dx$. As a consequence, $\phi^{a}$ 
can be shifted by an integer multiple of $2\pi$ by performing a 
topologically non-trivial gauge transformation which is single-valued 
in $SU(N)/Z_{N}$. As there are no fields in the theory which are charged 
under the center of the gauge group, it is consistent to divide out by all 
such gauge transformations. Thus we learn that each $\phi^{a}$ 
is a periodic variable with period $2\pi$.    
\paragraph{}
In addition to the $N$ periodic scalars $\phi^{a}$, the low energy 
theory on the Coulomb branch also includes $N$ massless photons $A^{a}_{i}$, 
with $\sum_{a=1}^{N}A^{a}_{i}=0$, 
where $i=1,2,3$ is a three-dimensional Lorentz index. 
In three dimensions, a massless abelian 
gauge field can be eliminated in favour of a scalar by a duality 
transformation. 
Thus we can eliminate the gauge fields 
$A^{a}_{i}$, in favour of $N$ scalars $\sigma^{a}$ subject to 
$\sum_{i=1}^{N}\sigma^{a}=0$. In particular, the 
fields $\sigma^{a}$ appear in the effective action as Lagrange multipliers 
for the Bianchi identity: 
${\cal L}_{\sigma}\sim \sigma\varepsilon^{ijk}\partial_{i}F_{jk}$.  
After integrating by parts, we obtain a term which only depends on 
$\partial_{i}\sigma^{a}$ not on $\sigma^{a}$ itself. 
However, integrating by parts also gives rise to a 
surface term in the action which is proportional to the magnetic 
charge. Specifically, the normalization of $\sigma^{a}$ is chosen 
so that the resulting term in the action is precisely 
$-i\sigma^{a}k^{a}$ where $k^{a}$ is the magnetic charge in the 
$U(1)$ subgroup corresponding to the Cartan generator $H^{a}$. 
The Dirac quantization condition implies that each 
magnetic charge is quantized in integer units: 
$k^{a}\in Z$, and so the path integral is invariant under 
shifts of the form 
$\sigma^{a}\rightarrow \sigma^{a}+ 2\pi n^{a}$ with $n^{a}\in Z$. 
Thus, like $\phi^{a}$, each $\sigma^{a}$ is a periodic variable with 
period $2\pi$. 
\paragraph{}
In the simplest case of gauge group   
$SU(2)$, we have two periodic scalars $\phi=\phi_{1}-\phi_{2}$ and 
$\sigma=\sigma_{1}-\sigma_{2}$.  After dividing out by the Weyl group 
of $SU(2)$, the classical Coulomb branch 
is $T^{2}/Z_{2}$ where $T^{2}$ is the two dimensional torus parametrized 
by $\phi$ and $\sigma$.    
The $Z_{2}$ gauge symmetry has four fixed points at $(\phi,\sigma)
=(0,0)$, $(\pi,0)$, $(0,\pi)$ and $(\pi,\pi)$. The fixed point at 
the origin $(\phi,\sigma)=(0,0)$ is special because, at least in the classical 
theory, the non-abelian gauge symmetry is restored at this point. This means,  
in particular, that there are additional massless degrees of freedom and 
a naive description in terms of the fields $\phi$ and $\sigma$ should 
break down. In fact, the three-dimensional duality transformation by which 
$\sigma$ was introduced is only possible when the low-energy theory 
is abelian and hence cannot be implemented at this point. 
We will find below that 
the origin continues to have a distinguished role in the 
quantum theory.  In contrast, the other three fixed points 
can be taken at face value as orbifold points of the $Z_{2}$ gauge symmetry: 
no additional massless degrees of freedom appear at these points. 
 \paragraph{}
 The low-effective energy effective action 
on the classical Coulomb branch of the $SU(2)$ theory 
is obtained by compactification of a  
four-dimensional $U(1)$ gauge theory with gauge coupling   
$g^{2}$ and vacuum angle $\theta$ \cite{SW3}. The 
resulting three-dimensional 
theory is a non-linear $\sigma$-model with target space $T^{2}/Z_{2}$. 
The bosonic terms in the action are,        
\begin{equation}
S_{\rm eff}=\frac{1}{16\pi R}\int\, d^{3}x\, \left(\frac{4\pi}{g^{2}}
\partial_{\mu}\phi\right)^{2}+ \left(\partial_{\mu}\sigma+\frac{\theta}{2\pi} 
\partial_{\mu}\phi\right)^{2}          
\label{leff2}
\end{equation}
This Lagrangian has a unique extension with ${\cal N}=1$ 
supersymmetry. To make this 
explicit we combine the real scalars $\phi$ and $\sigma$ to form 
a complex scalar $X=-i(\sigma+\tau\phi)$ where 
$\tau=4\pi i/g^{2}+\theta/2\pi$ is the usual complexified coupling. 
The action becomes, 
\begin{equation}
S_{\rm eff}=\frac{1}{16\pi R}\int\, d^{3}x\, \frac{1}{{\rm Im} \tau}
\,\partial_{\mu}X \partial_{\mu}\bar{X}          
\label{leff3}
\end{equation}    
This is the standard action for a non-linear $\sigma$-model with a 
complex torus $E$ as target. $E$ has periods $2\omega_{1}=2\pi i$ and 
$2\omega_{2}=2\pi i\tau$, and area $A=1/16\pi R$. 
The ${\cal N}=1$ supersymmetric 
extension of the action is obtained by promoting $X$ to a chiral 
superfield. The resulting Lagrangian can be written as a D-term 
in  ${\cal N}=1$ superspace, 
\begin{equation}
{\cal L}_{\rm eff}= \int\, d^{2}\theta d^{2}\bar{\theta} \, 
{\cal K}_{cl}[X,\bar{X}]
\label{lsuper1}       
\end{equation}    
with the K\"{a}hler potential 
${\cal K}_{cl}=X\bar{X}/(16\pi R {\rm Im} \tau)$. 
\paragraph{}
The above discussion generalizes easily to gauge group $SU(N)$ with 
$N\geq 2$. 
In this case we have $N$ chiral superfields 
$X^{a}=-i(\sigma^{a}+\tau\phi^{a})$. For a generic point 
on the classical Coulomb branch the unbroken gauge group is $U(1)^{N-1}$. 
On submanifolds where one or more of the $X^{a}$ coincide, a non-abelian 
gauge symmetry is restored. As for $SU(2)$, 
the scalar fields naturally take values on a complex torus $E$ with 
complex structure parameter $\tau$. We must also divide by Weyl gauge 
transformations which permute 
the $N$ Cartan generators. Thus the $U(N)$ Coulomb branch is the 
symmetric product ${\rm Sym}^{N}(E)= E^{N}/S^{N}$ where $S^{N}$ is 
the permutation group of $N$ objects. As above, we restrict to gauge group 
$SU(N)$ by imposing $\sum_{a=1}^{N}X^{a}=0$. The low-energy effective action 
on the Coulomb branch is therefore ${\cal N}=1$ supersymmetric 
$\sigma$-model with target $E^{N-1}/S^{N}$. The K\"{a}hler metric on the 
target manifold is obtained from the K\"{a}hler potential ${\cal K}_{cl}=
\sum_{a=1}^{N}X^{a}\bar{X}^{a}/(16\pi R{\rm Im}\,\tau)$. 
\paragraph{}
At a generic point on the classical Coulomb branch the chiral superfields 
$X^{a}$ are the only massless degrees of freedom. 
As discussed above, the only exceptional points are those at which a 
non-abelian subgroup of the gauge group 
is restored. 
Away from these points all other fields in the theory 
acquire masses via the Higgs mechanism. In this case we expect to be able 
integrate out the massive degrees of freedom and derive an effective action 
for the $X^{a}$. 
As we must preserve ${\cal N}=1$ supersymmetry, the 
most general possible action, including terms with at 
most two derivatives or four fermions is,      
\begin{equation}
{\cal L}_{\rm eff}= \int\, d^{2}\theta d^{2}\bar{\theta} \, 
{\cal K}[X^{a},\bar{X}^{a}] \,\, + \, \, \int\, d^{2}\theta\, 
{\cal W}(X^{a}) 
\,\, + \,\, \int \, d^{2}\bar{\theta}\, \bar{{\cal W}}(\bar{X}^{a}) 
\label{lsuper2}       
\end{equation}  
where ${\cal W}(X^{a})$ is a holomorphic superpotential which lifts 
all or part of the vacuum degeneracy. The supersymmetric vacua of the theory 
are then determined by the stationary points of the superpotential. 
In general, this description of the low-energy physics is self-consistent 
only if the resulting vacua are located away from the point $X=0$. 
\paragraph{}
In the classical 
theory the K\"{a}hler potential is given by 
${\cal K}_{cl}$ defined above and the superpotential vanishes identically. 
Quantum corrections will modify both the D-term and the 
F-term in the superspace action. In general the exact effective action 
will depend in a complicated way on the parameters of the 
theory, including $\tau$, $R$ and $m_{i}$, with $i=1,2,3$.  
As usual, corrections to the K\"{a}hler 
potential are relatively unconstrained and therefore hard to determine. 
In contrast, the superpotential is constrained by holomorphy in the 
fields $X^{a}$. By standard arguments \cite{S1}, the superpotential is also 
holomorphic in the complexified coupling $\tau$ and the masses $m_{i}$. 
In many similar situations in three and four dimensions, global symmetries 
provide further constraints on ${\cal W}$. The present case is somewhat 
different as, for generic values of the masses $m_{i}$, the theory has 
no unbroken global symmetries. However, as we discuss 
below, we have instead a different 
constraint which is special to theories with compact dimensions: each 
chiral superfield $X^{a}$ is a complex variable with two periods. 
In Section 3, the consequences of double-periodicity and holomorphy will 
be exploited to to determine the exact superpotential. However, 
as a preliminary,  we will first discuss the quantum corrections 
to the superpotential in the weak-coupling limit $\tau \rightarrow i\infty$. 
\section{Semiclassical Analysis}
\paragraph{}
In the weak coupling limit, the path integral is dominated 
by field configurations of minimum action in each topological sector. 
In the following discussion we will restrict our attention to the Euclidean 
theory with gauge group $SU(2)$. 
Gauge field configurations in the compactified theory are 
labelled by 
two distinct kinds of topological charge \cite{GPY}. 
The first is the Pontryagin 
number carried by instantons in four-dimensions, 
\begin{equation}
p=\frac{1}{8\pi^{2}}\int_{R^{3}\times S^{1}}{\rm Tr}
\left[ F\wedge F\right]\,  
\label{4dinumber}
\end{equation} 
The 4D instanton number appears in the microscopic action as the term 
$-ip\theta$. For each value of $p$, the action is minimized by 
configurations which solve the (anti-)self-dual Yang-Mills equation.  
An important feature of the compactified theory which differs 
from the theory on $R^{4}$ is that $p$ is not quantized in integer units.   
However, for each integer value of $p$, the theory has 
solutions with action $S_{cl}=8\pi^{2} |p|/g^{2}-ip\theta$. 
Thus we have $S_{cl}= 2\pi i p\tau$ 
for $p>0$ and $S_{cl}=-2\pi i p\bar{\tau}$ for $p<0$. When the instanton scale 
size is much less than the compactification radius, these solutions are 
close to their counterparts on $R^{4}$, thus we will call them 
four-dimensional (4D) instantons.     
\paragraph{}
In addition to the usual four-dimensional instantons, the compactified theory 
also has finite action configurations which carry three-dimensional 
magnetic charge,  
\begin{equation}
k=\frac{1}{8\pi}\int_{R^{3}} \vec{\nabla}\cdot  \vec{B} \in Z
\label{magcharge} 
\end{equation}
where $B_{i}=\varepsilon_{ijk}F^{jk}/2$ is the abelian magnetic 
field of the low-energy theory.   
A useful way to think about these configurations is 
to introduce a fictitious fifth dimension. As our starting point 
is a Euclidean theory with four space-like dimensions, we will interpret 
the new coordinate as time. With this interpretation, the Wilson 
lines play the role of the adjoint Higgs fields in the Bogomol'nyi 
equation and the theory has static BPS monopole solutions. 
As usual, quantization of 
magnetic charge can be understood either from Dirac's argument or 
by identifying $k$ with an element of the 
non-trivial homotopy group $\pi_{2}[SU(2)/U(1)]=Z$. 
As these configurations do not depend either on the time or 
on the coordinate 
of the compactified spatial dimension, we will refer to them as 
three-dimensional (3D) instantons. 
In addition to the integer-valued magnetic charge $k$, these 3D 
instantons also carry fractional Pontryagin number $p=\phi/2\pi$ \cite{LY}.  
The magnetic charge appears in the action as the term 
$-ik\sigma$ while, as above, 4D instanton number appears as the term 
$-ip\theta$. Including these term the action of a 3D instanton can be 
written as  
$S_{3D}=-i\tau\phi|k|+ik\sigma$. Thus we have $S_{3D}=kX$ for $k>0$, 
and $S_{3D}=k\bar{X}$ for $k<0$. 
\paragraph{}
As discussed in the previous section, ${\cal N}=1$ 
supersymmetry permits a non-zero holomorphic superpotential ${\cal W}(X)$. 
Although 
the superpotential is absent in the classical theory it can be generated 
by quantum effects \cite{AHW}. The leading semiclassical 
contribution of a 4D instanton (for $p>0$) is $\exp(-S_{4D})=
\exp(2\pi i p\tau)=q^{p}$ while that of a 3D instanton (for $k>0$) 
is $\exp(-S_{3D})=\exp(-kX)$.  
As these contributions are holomorphic in $\tau$ and $X$ respectively, 
both kinds of instantons potentially contribute to the superpotential 
at leading semiclassical order. 
In order to decide whether these contributions are non-zero, 
we need to examine the fermion zero modes of each type of instanton. 
In particular, the superpotential contains couplings which are bilinear 
in Weyl fermions of one four-dimensional chirality. 
To contribute, an instanton must therefore have exactly two zero modes 
of the same chirality. 
\paragraph{}
Before discussing
each type of instanton individualy, it is possible to make some comments 
which apply equally to 3D instantons, 4D instantons and also to some 
more exotic solutions which we will meet below. 
As both 3D and 4D instantons satisfy the self-dual 
Yang-Mills equation, both are invariant under supercharges of one 
four-dimensional chirality. However, 
the action of the remaining supercharges on the instanton is non-trivial and, 
by Goldstone's theorem, each generates a fermion zero mode. Thus in a theory 
with ${\cal N}=1$ supersymmetry (ie four supercharges), the existence 
of the required zero modes is guarenteed for both types of instanton. 
However, for any self-dual configuration, index theorems suggest the 
existence a much larger number of zero modes. Specifically, they suggest 
a number of zero 
modes which grows linearly with the topological charge. If these zero modes 
were really present, instanton contributions to the superpotential would 
vanish except, perhaps, in the sector of lowest topological charge. 
In fact, in a wide range of cases \cite{MO,DHKMV,DKM3D}, 
we know that almost all of these zero 
modes are lifted by quantum effects. Typically this is possible 
because the relevant index theorems do not take into account  
Yukawa couplings between fermions and the scalar fields of the theory. 
The only exceptions are zero modes whose existence is 
guarenteed by an unbroken symmetry of the theory. In particular, 
this is the case for the two zero modes corresponding to the action of 
the supercharges on the instanton.  
\paragraph{}
The 3D instantons which carry magnetic charge $k$ 
provide a concrete illustration of the general discussion 
given above.   
As these configurations are essentially BPS monopoles, the Callias index 
theorem predicts the existence of $2k$ zero modes for each species of 
massless adjoint Weyl fermion. 
It is convenient to start by considering the case 
when the three masses $m_{i}$ are set to zero.  In 
this case, 
we have an ${\cal N}=4$ supersymmetric theory with four massless species of 
Weyl fermion and we find a total of $8k$ zero modes. As we now have 
sixteen supercharges, eight 
of these modes are Goldstone modes for the half of the supersymmetry
algebra which acts non-trivially on self-dual configurations. 
These modes are protected 
from quantum corrections by supersymmetry and therfore remain as exact 
zero modes in the ${\cal N}=4$ quantum theory. 
The remaining modes are not protected by any symmetry and can 
thus be lifted in the quantum effects. In fact, at least 
in the three-dimensional limit, this lifting was demonstrated 
explicitly in \cite{DKM3D}. The lifting occurs in a by now familiar way: 
the classical instanton action acquires a four-fermion term which couples to 
the Riemann tensor on the monopole moduli space. Because of this 
effect, 3D instantons have eight exact zero modes 
and therefore contribute to an eight fermion 
vertex in the effective action for each value of $k$.  
If we break ${\cal N}=4$  
supersymmetry down to ${\cal N}=1$ by reintroducing non-zero masses  
for the three adjoint 
chiral multiplets, six of the eight zero modes are lifted and the instantons 
can contribute directly to the superpotential. The leading semiclassical 
contribution of 3D instantons of positive magnetic charge $k$, therefore 
has the form $m_{1}m_{2}m_{3}\exp(-kX)$. 
\paragraph{}
The form of the contribution of 3D instantons to the superpotential 
immediately presents us with a puzzle. In the classical theory, 
$\phi$ and $\sigma$ are both periodic variables with period $2\pi$. 
At weak coupling, we 
might expect the action to respect both these classical periodicities. 
This point will be discussed further below. For the moment it suffices 
to note that, as we have $X=-i(\sigma+ \tau\phi)$, 
the instanton-generated term in the superpotential 
is manifestly periodic in $\sigma$. However 
it is certainly not periodic in $\phi$: if we shift 
the Wilson line $\phi$ by  $2\pi$, $\exp(-X)$ picks up a factor of 
$q=\exp(2\pi i\tau)$, the factor associated with a  
4D instanton! In fact this reflects a phenomenon which has recently been 
studied in detail by Lee and Yi \cite{LY}. Recall that a single 
BPS monopole can 
be thought of as a radial 
kink in three-dimensional space which interpolates between a 
Coulomb phase vacuum at spatial infinity and the symmetric vacuum 
at the origin. 
Similarly the 3D instanton described above 
interpolates between a vacuum with non-zero 
Wilson line $\phi(r=\infty)=\phi$ 
at radial infinity and the symmetric vacuum with $\phi(r=0)=0$  
at the origin. However as $\phi$ is periodic, we could equally 
well choose a solution of the Bogomol'nyi equations which has 
$\phi(r=\infty)=\phi+2n\pi$ for any integer $n\geq 1$. 
The new configuration is just the 
standard Prasad-Sommerfeld solution 
appropriately rescaled to match the shifted value of the asymptotic Wilson 
line. These solutions have Pontryagin number $n+\phi/2\pi$ and 
Euclidean action, 
\begin{equation}
S^{(n)}_{+}=\frac{4\pi}{g^{2}}\left(\phi + 2\pi n\right) 
-i\sigma-i\theta\left(\phi+2\pi n\right)= X-2\pi i \tau n
\label{shift}
\end{equation}
for each integer $n\geq 1$. 
\paragraph{}
So far we have an infinite tower of 3D instantons, each with magnetic 
charge $k=1$, labelled by a positive integer $n$. The action of each 
member of the tower is related to that of the ordinary 3D 
instanton by the replacement $\phi\rightarrow \phi +2\pi n$. 
The idea now is that we can potentially obtain a superpotential which 
is periodic in $\phi$ by summing over sectors of different $n$.  
However, for this to have any chance of working, we also need to find 
a solution corresponding to the replacement 
$\phi \rightarrow \phi-2n\pi$ for each $n$. 
In fact the necessary solutions are also discussed in \cite{LY} and 
the surprise is that they have negative magnetic charge $k=-1$. Importantly 
these are solutions of the same Bogomol'nyi equation $B_{i}=+D_{i}\Phi$ 
as the ordinary $k=+1$ monopole, and thus are  
{\em not} related to the anti-monopole which satisfies $B_{i}=-D_{i}\Phi$. 
This possiblity arises  
because of the periodicity of the Wilson line 
which allows us to consider Higgs fields which have a positive value 
at the origin and decrease in the radial direction.    
By changing the sign of the radial derivative of $\Phi$ 
we then change the sign on the right-hand side of the Bogomol'nyi 
equation. In this way we obtain an infinite tower of $k=-1$ solutions 
with action,  
\begin{equation}
S^{(n)}_{-}=\frac{4\pi}{g^{2}}\left(2\pi n-\phi\right) 
-i\sigma-i\theta\left(2\pi n-\phi\right)= -X-2\pi i \tau n
\label{shift2}
\end{equation}
for each integer $n\geq 1$. The fact that these configurations have 
negative magnetic charge yet satisfy the 
Bogomol'nyi equation with positive sign is reflected in the fact that 
the action is holomorphic in $X$ rather than anti-holomorphic as would 
be the case for the action of an anti-monopole. 
\paragraph{}
The above discussion should generalize 
in a straightforward way to higher magnetic charge: in addition to the usual 
charge-$k$ 3D instanton with 
action $kX$ we find two infinite towers of solutions 
with action $kS^{(n)}_{+}$ and $kS^{(n)}_{-}$ respectively, $n\geq 1$,   
with magnetic charges $+k$ and $-k$. Also, the analysis of the 
fermion zero modes given above for the ordinary charge-$k$ 3D instanton 
solution applies equaly to each member in the tower: all zero modes not 
protected by supersymmetry should be lifted by quantum effects. Including 
the effects of non-zero masses, we find that configurations in each sector 
have only two exact zero modes and can therefore contribute to the 
superpotential. 
\paragraph{}
Finally, we still have to discuss the contribution of ordinary 4D instantons. 
As discussed in \cite{LY}, 
4D instantons have a very natural description in terms of 
the configurations described above. Consider for example a configuration 
corresponding to an ordinary 3D instanton with $k=+1$ together with the lowest 
member ($n=1$) of the $k=-1$ tower at large spatial separation. A priori 
this is only an approximate solution of the field equations. 
However, because of the sign flip in the Bogomol'nyi equation, 
the forces between the two constituent 
objects cancel and it is plausible that a corresponding exact solution 
of the self-dual Yang-Mills equations exists. 
As the action of this configuration is $2\pi i \tau$,  this 
solution is  
naturally identified with a single Yang-Mills instanton on 
$R^{3}\times S^{1}$. According to the Atiyah-Singer index theorem a 
single $SU(2)$ instanton has four fermion zero modes per species of 
massless Weyl fermion. In the presence of non-zero masses $m_{i}$, 
we have a single massless gluino and thus four zero modes. This 
counting agrees with the fact that the two constituent 3D instantons 
have two zero modes each. However, as above, only two zero modes are 
protected by supersymmetry and we expect the two remaining modes 
to be lifted. In the four-dimensional theory, the extra 
modes are generated by superconformal transformations in the instanton 
background and are protected if this symmetry remains unbroken. 
In the present case, even when the masses $m_{i}$ are zero, 
superconformal invariance is broken by compactification on 
$R^{3}\times S^{1}$. 
\paragraph{}
Similar considerations indicate that 4D instantons can contribute to 
the superpotential for each topological charge $p>0$.  
These configurations have 
magnetic charge zero and so they simply 
contribute a constant term proportional  
to $m_{1}m_{2}m_{3}q^{p}$. 
As only the derivatives 
of the superpotential with respect to  
the chiral field $X$ appear in the low energy effective action, 
one might wonder whether such a term has any significance. In fact 
these terms contribute directly to the Greens functions of chiral 
operators. For example, the gluino condensate is given by the standard 
relation, $\langle \lambda \lambda \rangle=\partial_{\tau}{\cal W}$, which is 
obtained by promoting $\tau$ to a background chiral superfield \cite{P}. 
\paragraph{}
Putting everything together we can write out the most general possible 
superpotential which can generated at leading semiclassical order, 
\begin{eqnarray}
{\cal W} & = & m_{1}m_{2}m_{3}\sum_{p=1}^{\infty}a_{p}q^{p}+ 
 m_{1}m_{2}m_{3}\sum_{k=1}^{\infty}  
 b_{k}\exp(-kX)  \nonumber \\ & & \qquad{} \qquad{} 
 \qquad{}  +  m_{1}m_{2}m_{3}\sum_{k=1}^{\infty}\,\sum_{n=1}^{\infty} 
 c_{k,n}q^{kn}\exp(-kX) +
 d_{k,n}q^{kn}\exp(+kX) \nonumber \\
 \label{expansion}
 \end{eqnarray}
In fact, we can immediately rule out any perturbative corrections to this 
result by noting that dependence on the vacuum angle 
$\theta$ cannot arise in perturbation theory. 
By holomorphy in $\tau$, any powers $g^{2}$ appearing in the 
superpotential are necessarily 
accompanied by $\theta$-dependence and are thus forbidden. 
This argument applies equally to 
perturbation theory in the vacuum sector or in an instanton background. 
The exact superpotential is therefore 
completely determined at leading semiclassical order.   
This means in particular that the 
dimensionless coefficients $a_{p}$, $b_{k}$, $c_{k,n}$ and $d_{k,n}$ 
do not depend on $X$ or on $\tau$. However, at this 
 point it appears that they can be arbitrary functions of the three 
 dimensionless parameters $m_{i}R$.
In the rest of the paper we present 
 arguments which determine the exact superpotential and, therefore,  
 the corresponding coefficients. In particular we 
  will discover that the superpotential is actually independent of $R$. 
\paragraph{}
 The discussion given above implicitly assumes that the quantum theory has a  
 regime in which the degrees of freedom corresponding to the 
 chiral superfield $X$ are weakly-coupled and a semiclassical analysis makes 
 sense. In such a regime the instanton corrections appearing on the RHS of 
 (\ref{expansion}) should be exponentially small. Apart from requiring 
  $g^{2}<<1$, this condition also constrains $X$.  
 In similar situations 
 in three or four dimensions, we could ensure weak coupling by going 
 to an asymptotic region of the moduli space where the scalar fields have 
 large expectation values. The present case is more subtle 
 because the Coulomb branch is a compact torus. 
 In particular, this is 
 reflected by the presence of terms of order $q^{k}\exp(+kX)$ 
 in (\ref{expansion}) which {\em grow} exponentially as $g^{2}\rightarrow 0$ 
 if $\phi>2\pi$. Similarly the terms of order $\exp(-kX)$ grow at 
 weak coupling for $\phi<0$. However, provided we choose a 
 Wilson line in the range $0<\phi <2\pi$, 
 then every term in (\ref{expansion}) 
 vanishes exponentially as $g^{2}\rightarrow 0$. In this region of field 
 space, the weak 
 coupling analysis given in this section is reliable. At first sight this 
 restriction on $\phi$, seems to contradict the fact that the classical 
 theory is periodic in this variable. As anticipated above, the resolution 
of this puzzle must be that the periodicity of 
${\cal W}$ in $\phi$ is only visible 
after summing over Lee-Yi copies and multi-instantons. A semiclassical 
expansion of the superpotential can then be performed for any value of 
$\phi\neq 2n\pi$ after an appropriate change of variables. 
In the next Section, this idea will be implemented explicitly.            
\section{The Exact Superpotential} 
  \paragraph{}
 The superpotential of the $SU(2)$ theory 
is a holomorphic function on the chiral superfield 
$X=-i(\sigma+\tau\phi)$. The weak coupling 
arguments given in the previous Section indicate that this function 
should respect the classical periodicity of the variables $\phi$ and $\sigma$. 
Thus, ${\cal W}(X)$ is a holomorphic function defined on the complex torus 
$E$. The superpotential should also be invariant under the Weyl group of 
$SU(2)$ which acts as $X\rightarrow -X$. As ${\cal W}(X)$ is a 
non-constant holomorphic 
function on a compact domain it must have a singularity somewhere.    
As usual such a singularity should correspond to a point at which 
our low-energy description of the theory breaks down due to the 
presence of extra light degrees of freedom. There is only one such point 
on the classical moduli space: the origin $X=0$ where non-abelian gauge 
symmetry is restored. Thus we will assume that ${\cal W}$ has an isolated 
singularity at this point and no other singularities. In fact we 
  will limit ourselves further by concentrating on the minimal case 
  where the singuarity is a pole, providing numerous checks that the 
  resulting solution is correct. At first sight, the presence of a 
  singularity at the origin in the quantum theory might seem to be 
  in conflict with our understanding of ${\cal N}=2$ theories in 
  four dimensions \cite{SW1,SW2}. 
  In such theories, the classical singularity at the 
  origin of the Coulomb branch is split by strong-coupling effects 
  into two or more singularities 
  where BPS monopoles or dyons become massless. However, 
in the present context, 
four-dimensional monopoles correspond to instantons and do not appear as 
states in the Hilbert space. Thus candidates for the massless states 
required to explain multiple singularities are absent. 
  \paragraph{}
  It follows from the above discussion that we are looking for a 
  meromorphic function on the torus $E$ which is even under 
  the action of the Weyl group. It is convenient to introduce 
  an explicit realization of the 
  torus $E$ as the quotient $C/\Gamma$ where $\Gamma$ is the lattice, 
  \begin{equation}
  \Gamma=\{z=2m\omega_{1}+2n\omega_{2}: (m,n) \in Z^{2}\}
  \label{lattice}
  \end{equation} 
  with $\omega_{2}/\omega_{1}=\tau$. Comparing with the results of 
  Section 1.2 we have $\omega_{1}=i\pi$.  
  A period parallelogram is any set of the 
  form, 
  \begin{equation} 
  D_{\Gamma}(z_{0})=\{z=z_{0}+2\mu \omega_{1}+ 2\nu \omega_{2}: 
  \mu, \nu\in [0,1)\}
  \label{pp}
  \end{equation}
  complex $z_{0}\in C$. 
  A meromorphic function on $E$ is the same as a meromorphic 
  doubly-periodic function on the complex plane with 
  periods $2\omega_{1}$ and $2\omega_{2}$. These are 
  known as elliptic functions and they are very highly constrained objects 
  (see for example \cite{EF}).  
  Some relevant facts are: 
  \paragraph{}
  {\bf 1} Elliptic functions are classified by their {\em order},  
  $\gamma$, which is 
  equal to the total number of poles in any period parallelogram 
  $D_{\Gamma}$ counting 
  by multiplicity (ie a double pole contributes $+2$ to $\gamma$).  
  \paragraph{}
  {\bf 2} An elliptic function attains an arbitrary complex value $\gamma$ 
  times in a period parallelogram counting by multiplicity. 
  Thus the total number of zeros in $D_{\Gamma}$ is also equal to $\gamma$. 
  \paragraph{}
  {\bf 3} Only the constant function has $\gamma=0$ and there are no functions 
  with $\gamma=1$. The basic example of a function with $\gamma=2$ 
  is the Weierstrass function,
  \begin{eqnarray} 
  {\cal P}(X:\Gamma) &= & \frac{1}{X^{2}}+ \sum_{W\in \Gamma -\{0\}}
  \left [ \frac{1}{(X-W)^{2}}-\frac{1}{W^{2}} \right]   \nonumber \\
  & = & \frac{1}{X^{2}} + \sum_{m,n\neq (0,0)}
  \left [ \frac{1}{(X-2m\omega_{1}-2n\omega_{2})^{2}}-\frac{1}{
 (2m\omega_{1}+2n\omega_{2})^{2}} \right]                  
 \label{weier}
 \end{eqnarray}
 ${\cal P}(X)$ has a double pole at $X=0$. At $\gamma=2$ there are also 
 the Jacobian elliptic functions which have two single poles in $D_{\Gamma}$. 
 \paragraph{}
 {\bf 4} The derivative of an elliptic function is an elliptic 
 function with the same periods. 
 In fact ${\cal P}'(X)$ is related to ${\cal P}(X)$ by the 
 differential equation, 
 \begin{eqnarray}
 \left( \frac{d{\cal P}}{dX}\right)^{2} &= & 4{\cal P}^{3}-g_{2}(\tau){\cal P}-
 g_{3}(\tau) 
 \nonumber \\ & = & ({\cal P}-e_{1}(\tau))({\cal P}-e_{2}(\tau))
 ({\cal P}-e_{3}(\tau))
 \label{de}
 \end{eqnarray}
 where
 \begin{eqnarray}
 g_{2}=60\sum_{(m,n)\neq (0,0)}
  \frac{1}{
 (2m\omega_{1}+2n\omega_{2})^{4}} & &  g_{3}=140\sum_{(m,n)\neq (0,0)}
  \frac{1}{
 (2m\omega_{1}+2n\omega_{2})^{6}}
 \label{inv}
 \end{eqnarray} 
 \paragraph{}
 {\bf 5} The Weierstrass function is even: ${\cal P}(X)={\cal P}(-X)$. 
 Any even elliptic function can be expressed as a rational function 
 of ${\cal P}(X)$. Correspondingly any odd 
 elliptic function can be expressed as 
 ${\cal P}'(X)$ times a rational function of ${\cal P}(X)$. 
 \paragraph{}
 As the Weyl group action is $X\rightarrow -X$, ${\cal W}(X)$ must be an even 
 elliptic  function and thus, from {\bf 5}, a rational function of 
  ${\cal P}(X)$. In fact, because the only singularity of the superpotential 
 is at the origin, ${\cal W}$ must be a polynomial in ${\cal P}(X)$. The 
 order, $\gamma_{{\cal W}}$, of ${\cal W}(X)$ is then twice the degree of 
 this polynomial. From ${\bf 4}$, we also learn that 
 the derivative of ${\cal W}$ is an odd elliptic function of order 
 $\gamma_{\cal W}+1$. To pin down the exact function we seek, 
 we must provide one more piece of physics input. 
 Recall that the supersymmetric vacua of the theory 
 correspond to the zeros of $\partial {\cal W}/\partial X$. 
 Simple zeros correspond to massive vacua while a higher order zero 
 yields a vacuum with massless particles. Following \cite{SW3}, 
we will assume that the physics varies smoothly with $R$ and correspondingly 
we expect the number and type of vacua to be the same as in the 
 four-dimensional theory. According to the arguments in Section 1.1, 
 the four-dimensional theory with gauge group $SU(2)$ and three massive 
 chiral multiplets has three vacua: two coming from the confining phase 
 where the low energy theory is ${\cal N}=1$ SYM and 
 an extra vacuum in which the gauge group is completely broken. 
 In each of these vacua, the theory has a mass gap. Three massive vacua 
 means that $\gamma_{\cal W}+1=3$. Finally, as there is only one 
 elliptic function of order two with a double pole at the origin        
 we have the result\footnote{Two dimensional Landau-Ginzburg theories 
 with this superpotential have been considered before. See \cite{CV} 
 and references therein.}, 
 \begin{equation} 
 {\cal W}(X)=m_{1}m_{2}m_{3}\left({\cal P}(X)+C(\tau) \right)
 \label{result}
 \end{equation}
Importantly, we have obtained a superpotential which is independent of $R$.  
Strictly speaking we have not determined the overall normalization of 
(\ref{result}). This will ultimately be determined (and shown to 
be independent of $R$) in the next Section. Note also that, so far, 
 we have only determined the superpotential 
up to the addititive constant $C(\tau)$. This ambiguity 
will be discussed further below. 
 \paragraph{}
 From the differential equation (\ref{de}) we see that, as required, 
 $\partial {\cal W}/\partial X$ has three simple zeros and that the 
 corresponding critical values ${\cal W}$ are $e_{1}(\tau)$, $e_{2}(\tau)$ 
 and $e_{3}(\tau)$. In fact these values are attained at the 
 half-lattice points $X=\omega_{1}=i\pi$ $X=\omega_{1}+\omega_{2}=
 i\pi(\tau+1)$ and $X=\omega_{2}=i\pi\tau$ respectively. These are three 
 of the four fixed points of the Weyl group discussed in Section 1.2 above. 
 In the remainder 
 of this section we highlight several interesting features of this result: 
 \paragraph{}
 {\bf Modular properties:} The lattice $\Gamma$ is 
 invariant under modular transformations acting on the 
 half-periods as $\omega_{2}\rightarrow a\omega_{2}+b\omega_{1}$  and 
 $\omega_{1}\rightarrow  c\omega_{2}+d\omega_{1}$ with $a,b,c,d\in Z$ 
 and $ad-bc=1$.  As the function ${\cal P}(X)$
 only depends on the choice of the lattice $\Gamma$ it is invariant under 
 these transformations. Recalling that 
 $\tau=4\pi i/g^{2}+\theta/2\pi= \omega_{2}/\omega_{1}$ we see 
 that these transformations correspond to the familiar action of S-duality 
 on the complexified coupling of the ${\cal N}=4$ theory, namely 
 $\tau\rightarrow (a\tau+b)/(c\tau+d)$. In fact this is not quite the 
 same as the action on the periods defined above because we have  
 $\omega_{1}=i\pi$ and $\omega_{2}=i\pi\tau$ so only $\omega_{2}$ 
 transforms. S-duality corresponds to a modular transformation 
 of the lattice accompanied by a rescaling by a factor of 
 $(c\omega_{2}+d\omega_{1})$ in order to preserve $\omega_{1}=i\pi$. 
 Taking this into account we discover that the superpotential ${\cal W}$ 
 transforms with modular weight two under S-duality. 
  This means that the F-term effective Lagrangian of the theory is 
 invariant under S-duality if we assign the superspace measure $d^{2}\theta$ 
 modular weight $-2$. 
 \paragraph{}
 It is also worth noting that {\em any} non-zero $C(\tau)$ would 
 spoil these modular 
 properties, or at least introduce unphysical poles for some 
 values of $\tau$. The reason for this is as follows: 
 while  ${\cal W}$, as function of $X$ and $\tau$,  
 is free to transform with modular weight two, a holomorphic function of 
 $\tau$ alone, with no poles in the fundamental domain of $SL(2,Z)$, cannot. 
  If it were to do so it would be a modular form of 
 weight two of which there are none \cite{KOB}. 
 Thus modularity of the superpotential uniquely picks $C(\tau)=0$. The 
 action of S-duality on the critical points of the potential is also  
 interesting. As above, ${\cal W}$ transforms with modular weight two, but its 
 critical values $e_{i}(\tau)$, as functions of $\tau$ alone, cannot. 
 The modular properties of the $e_{i}$ were described 
 in \cite{SW2}. Each is modular only with respect to a certain subgroup 
 of the modular group. Under the rest of the modular transformations, 
 the $e_{i}$ transform into each other. This means that S-duality 
 transformations permute the three vacua. This is very natural in the 
 light of the interpretation of these vacua given in \cite{SW2}. 
 In particular the three vacua are associated with the condensation 
 of elementary quanta, monopoles and dyons respectively. As the S-duality 
 of the ${\cal N}=4$ theory permutes these BPS states it is reasonable that 
 it should do the same to the corresponding vacua of the ${\cal N}=1$ theory. 
 \paragraph{}
 {\bf Semiclassical expansion:} In Section 2, 
 we argued that the exact superpotential should have a very 
 specific form at weak coupling with semiclassical contributions from 
 4D instantons, 3D instantons together with 
 two infinite towers of Lee-Yi copies. We can now expand our proposal for the 
 exact superpotential in the semiclassical limit and compare it with 
 these expectations. Expanding the Weierstrass function for 
 $\tau\rightarrow i\infty$ we obtain, 
\begin{eqnarray}
{\cal W} & = & m_{1}m_{2}m_{3}\left[\frac{1}{12}E_{2}(\tau) + \frac{1}{4} 
\sum_{n=-\infty}^{+\infty} \frac{1}{\sinh^{2}
\left(\frac{X+2\pi in\tau}{2}\right)}\right] \nonumber \\
& =& m_{1}m_{2}m_{3}\sum_{k=1}^{\infty}  
 k\exp(-kX) +  m_{1}m_{2}m_{3}\sum_{k=1}^{\infty}\,\sum_{n=1}^{\infty} 
 kq^{kn}\left[\exp(-kX) + \exp(+kX)-2\right]\ \nonumber \\
 \label{expansion2}
 \end{eqnarray}
 where $E_{2}(\tau)$ is the regulated Eisenstein series of modular weight 
 two \cite{KOB}. 
Comparing with (\ref{expansion}) we deduce that $a_{p}=\sum_{d|p}d$ and 
$b_{k}=c_{k,n}=d_{k,n}=k$. In hindsight, we could have deduced the 
$n$-independence of the coefficients $c_{k,n}$ and $d_{k,n}$ simply 
by requiring $2\pi$-periodicity of ${\cal W}$ in the 
Wilson line $\phi$. These predictions 
could be tested against first-principles field theory calculations.            
\paragraph{}
{\bf Two limiting cases:}
We will now consider the limit in which we obtain a 
three dimensional gauge theory. This corresponds to 
taking the limit $R\rightarrow 0$, 
$\tau\rightarrow i\infty$ while keeping the three dimensional gauge coupling 
$e^{2}=g^{2}/R$ fixed. 
In this limit, the period of the torus associated with 
the Wilson line goes to infinity while the period corresponding to the 
dual photon $\sigma$ stays fixed. Hence, both the 4D 
instantons and the towers of Lee-Yi copies have infinite 
action in this limit and 
the corresponding contributions to ${\cal W}$ vanish giving,  
\begin{equation} 
{\cal W}=m_{1}m_{2}m_{3}\frac{1}{\sinh^{2}\left(\frac{X}{2}\right)}
\label{3d}
\end{equation}
This is a prediction for the superpotential of a theory with 
${\cal N}=1$ supersymmetry which can be obtained as a massive
deformation of the three-dimensional $SO(8)$ invariant conformal field theory 
with sixteen supercharges. 
Note that the superpotential has a double pole at the $SO(8)$ invariant 
point $X=0$ where the Coulomb branch description breaks down. The theory 
has a single supersymmetric vacuum at $X=i\pi$ which corresponds to an 
orbifold fixed points at which the corresponding theory with 
sixteen supercharges is believed to be free \cite{S2}. 
In the weak coupling limit we can re-expand the 
 superpotential and 
 see that it includes an infinite series of 3D instanton corrections
 closely related to the 3D instanton series summed in \cite{DKM3D} 
 for the theory with sixteen supercharges. 
\paragraph{}
Another interesting limit is obtained by keeping $R$ finite but decoupling 
the three chiral multiplets. This gives four-dimensional 
${\cal N}=1$ SYM theory with gauge group $SU(2)$ 
compactified on a circle of radius $R$. Thus we take the limit 
 $m_{i}\rightarrow\infty$ and $\tau\rightarrow i\infty$ with 
 $\Lambda=(m_{1}m_{2}m_{3})^{\frac{1}{3}}q^{\frac{1}{6}}$ held fixed. 
 Naively, the superpotential appears to be divergent in this case.    
 In fact to obtain the correct answer we need to redefine our fields 
 according to $X=Y-\pi i \tau$. In terms of $Y$, we see that precisely 
 two terms in (\ref{expansion2}) survive in this limit: 
 the contribution from the 3D instanton with magnetic charge $k=1$, 
 which goes like $q^{\frac{1}{2}}\exp(-Y)$, and the $n=1$ 
 term in the tower of states with magnetic charge $-1$ which goes like 
 $q^{\frac{1}{2}}\exp(+Y)$. Thus we find, 
 \begin{equation}
  {\cal W}=\Lambda^{3}\left[\exp(-Y)+\exp(+Y)\right]    
 \label{sym}
 \end{equation}
 This agrees with the result obtained in \cite{SW3,3DG}, which has also 
 recently been 
 derived from first principles in \cite{HKM}. The fact that only two 
topological sectors contribute in this limit 
is due to a global symmetry of the 
minimal ${\cal N}=1$ theory which is broken explicitly in the 
full theory by couplings of the adjoint matter multiplets. 
The theory has two 
 supersymmetric vacua, located at $Y=\pm i\pi$, which agrees with the 
 Witten index of ${\cal N}=1$ SUSY Yang-Mills with gauge group $SU(2)$. 
 \paragraph{} 
 Note that the two limits considered above are complimentary in a 
 certain sense. 
 Specifically, we started from a Coulomb branch $T^{2}/Z_{2}$ with four 
 $Z_{2}$ fixed points of which three correspond to massive vacua. Two of these 
 vacua correspond to the confining phase of the four-dimensional theory while 
 one vacuum corresponds to the Higgs phase. The remaining 
  fixed point is a singular point of the superpotential. In the weak 
 coupling limit, $\tau \rightarrow \infty$, one period of the torus diverges 
 and the Coulomb branch becomes an infinite cylinder. However, which points 
 on the torus are sent to infinity depends on how 
 $X$, $m_{i}$ and $R$ scale in 
 this limit. The three-dimensional limit considered above 
 corresponds to blowing up the region 
 which contains the Higgs vacuum and the singularity. This sends 
 the two confinement 
 vacua to infinity. In contrast, the limit which yields ${\cal N}=1$ 
 SYM on $R^{3}\times S^{1}$ corresponds to blowing up the 
 neighbourhood of the two confining vacua which sends the Higgs vacuum and 
 the singular point to infinity. 
\paragraph{}
{\bf Generalisation to gauge group $SU(N)$:} The results 
described above have a fairly obvious generalization
  to gauge group $SU(N)$. In this section, rather than giving 
  detailed arguments, I will motivate an educated guess for the   
  exact superpotential. In subsequent sections I will present a detailed 
  check that the answer is the correct one. 
  \paragraph{}     
  For gauge group $SU(N)$, we have $N$ chiral superfields 
 $X^{a}$ $a=1,2,\ldots,N$ with $\sum_{a=1}^{N} X^{a}=0$ and the classical 
 Coulomb branch is $E^{N-1}/S^{N}$. The low energy gauge group 
 is $U(1)^{N}$ and thus we have $N-1$ species of 3D instantons which correspond to magnetic monopoles associated with each simple root of $SU(N)$. 
 For $a=1,\dots N-1$ these 
 contribute terms of order 
 $m_{1}m_{2}m_{3}\exp(X^{a}-X^{a+1})$ to the superpotential 
 \cite{DB,3DG}. Each of the 
 fundamental monopoles will have a towers of Lee-Yi copies 
 as well as a series of multi-monopole corrections which must be summed over.  
 Applying the arguments of the previous sections to constrain the 
 the holomorphic dependence on each of the complex 
 variables $X^{a}-X^{a+1}$, we can argue that the summation must 
 convert each exponential to the corresponding doubly periodic function 
 ${\cal P}(X^{a}-X^{a+1})$. 
 \paragraph{}
 However there is also a new feature which 
 enters for $N>2$. In particular we must also consider the contribution 
 of monoples embedded in $SU(2)$ 
 subgroups which correspond to non-simple roots of $SU(N)$. The relevant 
 configurations are ones which include fundamental monopoles charged 
 under two different magnetic $U(1)$'s. Index theorems indicate 
 that these configurations have more 
 fermion zero modes than a fundamental monopole. If this were true these 
 topological sectors would not contribute to the superpotential.   
 However, once again, the additional zero modes 
 are not protected by any symmetries and therefore will generically be lifted
 by quantum effects. This effect has been explicity demonstrated in 
 the three-dimensional theory with sixteen supercharges by 
 Fraser and Tong \cite{FT}. This suggests that we 
 should find contributions to the superpotential 
 which go like $m_{1}m_{2}m_{3}\exp(X^{a}-X^{b})$ for all $a \neq b$. 
 As above, summation over magnetic 
 charges and the towers of copies should promote each exponential to the  
 doubly periodic function ${\cal P}(X^{a}-X^{b})$. 
 Finally the results of \cite{FT} suggest that, after taking into account 
 the additional sectors, we should find complete 
 democracy among the roots of $SU(N)$ with no set of simple roots playing 
 a prefered role. Thus the 3D instanton corresponding to each root 
 should contribute to ${\cal W}$ with the same overall coefficent. 
 The proposal for the exact superpotential is therefore, 
 \begin{equation}
 {\cal W}(X)=m_{1}m_{2}m_{3}\sum_{a>b}{\cal P}(X_{a}-X_{b})  
 \label{sun}  
 \end{equation}
 where, taking 
 the action of the Weyl group into account, we have restricted 
 the range of the sum to $a>b$ to avoid double-counting. 
 \paragraph{}
 A detailed test of this proposal for arbitrary $N$ will be given in 
 Section 5. 
 However, we can immediately provide a check on the result in the simplest 
 non-trivial case of gauge group $SU(3)$. Recall that the vacua of the 
 corresponding 
 four dimensional theory are in one to one correspondence with the 
 three-dimensional representations of $SU(2)$. 
 According to the discussion in Section 1.1, the 
 $SU(3)$ theory has one Higgs phase vacuum corresponding to the irreducible 
 representation, three confining vacua corresponding to 
 the trivial representation and a 
 single Coulomb phase vacuum corresponding to the three-dimensional 
 representation of $SU(2)$ obtained by adding the trivial and fundamental 
 representations. As our analysis holds for all $R$, each  
 of these vacua should show up as a stationary point 
 of the proposed superpotential. 
 Thus, 
 \begin{equation}
 {\cal W}=m_{1}m_{2}m_{3}\left({\cal P}(X_{1}-X_{2}) + {\cal P}(X_{2}-X_{3})
  +{\cal P}(X_{3}-X_{1})\right)    
 \label{su3}
 \end{equation} 
 should have five stationary points, one of them massless. This means 
  that the coupled equations
  $\partial {\cal W}/\partial X_{a}$, $a=1,2,3$ must therefore have five 
  solutions modulo lattice translations and the action of the 
  Weyl group. Note that the Hessian matrix $H=\partial^{2} 
  {\cal W}/\partial X_{a}\partial X_{b}$, will always have one zero eigenvalue 
  corresponding to shifts in $X_{1}+X_{2}+X_{3}$ which is projected out 
  the condition  $X_{1}+X_{2}+X_{3}=0$ for gauge group $SU(N)$. In the four 
  predicted massive 
  vacua there must be no other zero modes. In contrast, in the predicted 
  massless vacuum, $H$ 
  should have at least one additional zero eigenvalue.     
\paragraph{}
 To check this we must solve the equations, 
 \begin{eqnarray}
 {\cal P}'(X_{1}-X_{2})- {\cal P}'(X_{3}-X_{1}) & =& 0 \nonumber \\
 {\cal P}'(X_{2}-X_{3})- {\cal P}'(X_{1}-X_{2}) & =& 0 \nonumber \\
 {\cal P}'(X_{3}-X_{1})- {\cal P}'(X_{2}-X_{3}) & =& 0 \
 \label{su3eq}
 \end{eqnarray}
 or simply ${\cal P}'(X_{1}-X_{2})= 
 {\cal P}'(X_{2}-X_{3}) ={\cal P}'(X_{3}-X_{1})$. Solutions should be counted 
 modulo lattice translations and permutations of 
 $X_{1}$, $X_{2}$ and $X_{3}$. We must also impose the constraint that 
 $X_{1}+X_{2}+X_{3}$ should vanish up to a lattice translation. 
 Direct calcuation 
 shows that the required zeros of $\partial {\cal W}$ are located at, 
 \begin{eqnarray}
 (X_{1},X_{2},X_{3}) & =& \left(0,\frac{2\omega_{1}}{3},
\frac{4\omega_{1}}{3} \right) 
 \nonumber  \\ 
 & & \left(0,\frac{2\omega_{2}}{3},\frac{4\omega_{2}}{3} \right) \nonumber  \\ 
 & & \left(0,\frac{2}{3}(\omega_{1}+\omega_{2}), 
\frac{4}{3}(\omega_{1}+\omega_{2})\right) \nonumber \\
  & & \left(0,\frac{2}{3}(\omega_{1}+2\omega_{2}), 
\frac{2}{3}(2\omega_{1}+\omega_{2})\right) \nonumber \\
  & &  \left(\omega_{1},\omega_{2},\omega_{1}+\omega_{2}\right) 
   \label{solns}
   \end{eqnarray}
   We find that, for generic values of $\tau$, the 
  Hessian matrix has only one zero eigenvalue at the first 
  four solutions in (\ref{solns}). It is easily checked that $H$ has one 
  extra zero eigenvalue at the fifth solution listed above. 
  Thus we find exact agreement 
  with the vacuum structure of the four-dimensional theory. It is 
  worth noting that this test would be failed by a superpotential which 
  included only the contributions from 3D instantons corresponding to simple 
  roots of $SU(3)$. Specifically the four massive vacua would be absent 
  in this case.               
    
 \section{Soft Breaking of ${\cal N}=2$ SUSY}
\paragraph{}
In the previous sections we analysed the vacuum structure of  
${\cal N}=1$ theories with with non-zero masses $m_{1}$, $m_{2}$ and $m_{3}$ 
in terms of softly-broken ${\cal N}=4$ supersymmetry. In this Section, 
based on the ideas of \cite{SW3}, we study the same theories 
from a different, and much more powerful, perspective.  If we set  
$m_{1}=0$ and $m_{2}=m_{3}=M$, the $SU(N)$ gauge theory studied 
above reduces to ${\cal N}=2$ supersymmetric Yang-Mills theory with 
a single adjoint hypermultiplet of mass $M$. 
In four dimensions, the exact Coulomb branch of the $SU(2)$ 
theory is governed by an elliptic curve as described 
by Seiberg and Witten \cite{SW2} for gauge group 
$SU(2)$. The corresponding curve for gauge group $SU(N)$, with 
$N\geq 2$, was given subsequently by Donagi and Witten \cite{DW}. 
The four-dimensional versions of the ${\cal N}=1$ theories studied above  
can then be realized by re-introducing a non-zero mass $m_{1}=\epsilon$ 
for the remaining massless chiral multiplet. In this approach, there is 
an elegant correspondence between the massive
vacua of the ${\cal N}=1$ theory and the singular points on the 
Coulomb branch of the underlying ${\cal N}=2$ theory. In fact, in the next 
section, we will show that the stationary points of our proposed 
superpotential reproduce this correspondence exactly.  
\paragraph{}
In a parallel 
development, Seiberg and Witten have also described how this analysis 
can be extended to apply to ${\cal N}=2$ theories on $R^{3}\times S^{1}$. 
We will begin this Section by 
reviewing the key points of their analysis in 
the simplest case of ${\cal N}=2$ supersymmetric Yang-Mills theory 
with gauge group $SU(2)$. The discussion will be very brief and we refer 
the reader to \cite{SW3} for further details. 
After this review, we will extend this analysis to the mass 
deformed ${\cal N}=4$ theories discussed in the previous sections. 
We will find that this yields an extremely simple derivation of the 
main results of this paper.   
\paragraph{} 
After compactification to $D=3$ on a circle of radius $R$, ${\cal N}=2$ 
SUSY Yang-Mills with gauge group $SU(2)$ 
has a Coulomb branch ${\cal M}$ of complex dimension two. One 
complex dimension corresponds to the adjoint scalar, 
$u=\langle{\rm Tr}\Phi^{2}\rangle$, 
which parametrizes the Coulomb branch of the four-dimensional theory. 
The other dimension is essentially the complex superfield $X$ introduced 
above, whose real and imaginary parts (for $\theta=0$) are the Wilson line 
and dual photon respectively. As the gauge fields have been eliminated 
in favour of scalars, the low-energy effective action is a 
three-dimensional non-linear $\sigma$-model with target ${\cal M}$. 
In order to preserve ${\cal N}=2$ SUSY (in the four-dimensional convention), 
${\cal M}$ must be hyper-K\"{a}hler. In general the 
hyper-K\"{a}hler metric will depend on $R$ in a complicated way. However, 
the key result of 
\cite{SW3} is that ${\cal M}$ has a distinguished complex structure which 
is independent of $R$. In particular, in terms of coordinates which are 
holomorphic with respect to the distinguished complex structure, 
 ${\cal M}$ is defined by the complex equation, 
\begin{equation}
y^{2}=x^{3}-x^{2}u+\Lambda^{4}x
\label{ec}
\end{equation}
where $\Lambda$ is the dynamical scale of the four-dimensional theory. 
This is exactly the same equation as that of the elliptic curve 
which governs the Coulomb branch of the four-dimensional theory. 
Indeed, near four-dimensions, this can be interepreted as an 
elliptic fibration of the $u$-plane, with the Seiberg-Witten 
curve of the $D=4$ theory as the fibre.    
\paragraph{}
The distinguished complex structure picks out an ${\cal N}=1$ subalgebra 
of the ${\cal N}=2$ supersymmetry algebra. In simple terms this just 
corresponds to taking the holomorphic coordinates of (\ref{ec}) and promoting 
them to ${\cal N}=1$ chiral superfields. As we will eventually 
break ${\cal N}=2$ supersymmetry down to this ${\cal N}=1$ subalgebra 
it is useful to determine the low-energy effective Lagrangian in a manifestly 
${\cal N}=1$ supersymmetric form. Fortunately, to determine the vacuum 
structure we will not need to know the whole effective action, only the  
F-term part.  Of course, if we knew the K\"{a}hler 
potential for the hyper-K\"{a}hler metric in terms 
of two unconstrained chiral superfields we could, in principle, write 
down the ${\cal N}=2$ $\sigma$-model action as a D-term in ${\cal N}=1$ 
superspace. In this case there would be no F-term. In fact, we 
have instead a description of the target space involving three chiral 
superfields, $x$, $y$ and $u$, subject to the holomorphic 
constraint (\ref{ec}). 
The constraint can be imposed by introducing a fourth chiral superfield 
$\lambda$ as a Lagrange multipler and considering superpotential 
\begin{equation} 
{\cal W}=\lambda\left(y^{2}-x^{3}+x^{2}u-\Lambda^{4}x\right) 
\label{lagr}
\end{equation}
\paragraph{}
In order to break ${\cal N}=2$ SUSY down to the ${\cal N}=1$ subalgebra 
which is manifest in (\ref{lagr}), we now introduce a superspace mass term, 
$\Delta{\cal W}=\epsilon u$, 
for the adjoint chiral multiplet. Making the convenient 
change of variables, $x-u=v$, $x=\Lambda^{4}\tilde{x}$ and 
$y=\Lambda^{4} \tilde{y}$ the superpotential becomes, 
\begin{equation} 
{\cal W}+ \Delta{\cal W}
=\lambda\Lambda^{8} \left(\tilde{y}^{2}-\tilde{x}^{3}+\tilde{x}^{2}u-
\Lambda^{4}\tilde{x}\right) +
\epsilon\left(\Lambda^{4}\tilde{x}-v\right)
\label{lagr2}
\end{equation}    
We can now solve the F-term equations, 
\begin{equation}
\frac{\partial {\cal W}}{\partial \lambda}=
\frac{\partial {\cal W}}{\partial \tilde{y}}= 
\frac{\partial {\cal W}}{\partial v}=0
\label{fte}
\end{equation}
with $\tilde{y}=0$, $\lambda=-\epsilon/\Lambda^{8}\tilde{x}^{2}$ and 
$v=-1/\tilde{x}$. 
On integrating out $\tilde{y}$, $\lambda$ and $v$ we obtain,
\begin{equation} 
{\cal W}=\epsilon\left(\Lambda^{4}\tilde{x}+ \frac{1}{\tilde{x}}\right)=
\epsilon\Lambda^{2}\left(\exp(-Y)+\exp(+Y)\right)
\label{n=2answer}
\end{equation} 
where we have mapped the complex $\tilde{x}$-plane to the cylinder 
parameterized 
by the dimensionless variable $Y$ with $\tilde{x}\Lambda^{2}=\exp(-Y)$. Thus 
we have reproduced the superpotential given in \cite{SW3,3DG} 
for ${\cal N}=1$ SYM theory compactified to three dimensions. 
Importantly the resulting superpotential is independent of $R$. 
This is an immediate consequence of Seiberg and Witten's proposal    
that distinguished complex structure is independent of $R$.  
\paragraph{}
Now we will follow precisely the same steps starting from the 
complex curve which describes mass-deformed ${\cal N}=4$  supersymmetric 
Yang-Mills theory with gauge group $SU(2)$. In terms of ${\cal N}=2$ 
supermultiplets, this model differs from the one considered above by 
the presence of an additional adjoint hypermultiplet of mass $M$. Although 
the Coulomb branch of this theory can be described in terms of a 
single elliptic curve, we will choose an alternative 
representation which generalizes more readily to $SU(N)$ with $N>2$. 
According to Donagi and Witten \cite{DW}, 
the Coulomb branch of the $SU(2)$ theory 
is determined by the pair of equations which can be written as, 
\begin{eqnarray}
y^{2}=F(x) &= & 4x^{3}-xg_{2}(\tau)-g_{3}(\tau) \nonumber \\
   & = & 4(x-e_{1}(\tau))(x-e_{2}(\tau))(x-e_{3}(\tau)) 
\label{plain}
\end{eqnarray}
and 
\begin{equation}
u=\frac{P^{2}}{2} +M^{2}x
\label{simple}
\end{equation}
The first equation describes a complex torus E with complex structure 
parameter $\tau$ and a single puncture corresponding to the point 
$x=y=\infty$ on the curve.  This is known as the `bare' spectral curve and, 
on its own, it describes an ${\cal N}=4$ theory with gauge group $U(1)$.  
The second equation, which is particularly simple for gauge group 
$SU(2)$, encodes the extra dependence of the $SU(2)$ theory 
on mass-parameters and VEVs. 
To be more precise we should identify 
points which are related by the action of $\alpha\beta$ 
\begin{eqnarray}
\alpha:\, y \rightarrow -y &\qquad{} \qquad{} & \beta:\, P \rightarrow -P
\label{ab}
\end{eqnarray}
\paragraph{}
Imitating the logic applied above for the theory without matter we expect 
the Coulomb branch of the four-dimensional theory compactified on a circle 
down to three dimensions to be a hyper-K\"{a}hler manifold ${\cal M}$ of 
complex 
dimension two with a distinguished complex structure specified by equations 
(\ref{plain}) and (\ref{simple}). The low energy effective action should 
be a three-dimensional non-linear $\sigma$-model with target ${\cal M}$. 
The $\sigma$-model Lagrangian can be written in ${\cal N}=1$ superspace 
in terms of chiral fields corresponding to the holomorphic coordinates 
$x$, $y$, $u$ and $t$ together with two Lagrange multipliers $\lambda_{1}$ 
and $\lambda_{2}$. Thus,  including the ${\cal N}=2$ breaking mass term, 
we start from the superpotential, 
\begin{equation}
{\cal W}=\lambda_{1}(y^{2}-F(x))+\lambda_{2}(\frac{P^{2}}{2}+M^{2}x-u)+ 
\epsilon u
 \label{lam12}
\end{equation}
The equations of motion for $\lambda_{1}$ and $\lambda_{2}$ are equations 
(\ref{plain}) and (\ref{simple}) respectively. Stationarizing the 
superpotential with respect to $x$ and $u$ gives the additional 
equations, $\lambda_{1}F'(x)=-M^{2}\lambda_{2}$, $\lambda_{2}=-\epsilon$. 
Together with Eqn (\ref{simple}), these 
two relations can be used to eliminate $\lambda_{1}$ $\lambda_{2}$ and $u$ 
in favour of $x$ and $P$. Substituting for these variables, the superpotential 
reduces to ${\cal W}=\epsilon u=\epsilon(P^{2}/2+M^{2}x)$. 
However we still have to impose 
equation (\ref{plain}) which constrains $x$ in terms of $y$. As mentioned 
above, Eqn (\ref{plain}) describes a complex torus in terms of two complex 
variables $x$ and $y$. We can now change variables to a single unconstrained 
complex variable $X$, which parametrizes the same torus, by setting 
$x={\cal P}(X)$ and $y={\cal P}'(X)$. Equation (\ref{plain}) is obeyed by 
virtue of the differential equation (\ref{de}) of Section 4. In terms 
of $X$ the superpotential now reads,
\begin{equation}
{\cal W}=\epsilon\left(\frac{P^{2}}{2}+M^{2}{\cal P}(X) \right)
\label{answer}
\end{equation}
and thus, setting $P$ to zero by it equation of motion, 
we recover the result of the previous section.   
As in the previous example, the superpotential is independent of $R$ 
simply because of the $R$-independence of the distinguished 
complex structure. This analysis also resolves the puzzle raised in the 
previous section concerning the singularity at the origin of the Coulomb 
branch of the ${\cal N}=1$ theory. The origin of the ${\cal N}=1$ Coulomb 
branch at $X=0$, corresponds to the point 
$x=y=\infty$ on the bare spectral curve. Thus 
the singularity is projected to infinity and does not appear on the 
Coulomb branch of the corresponding ${\cal N}=2$ theory.  
\paragraph{}
We now turn to the more challenging case of gauge group $SU(N)$. 
The Donagi-Witten solution of the $D=4$ 
theory is given in terms of two complex equations. The first is the 
bare spectral curve which appeared above in the $SU(2)$ case.       
\begin{eqnarray}
y^{2} & = &  4x^{3}-xg_{2}(\tau)-g_{3}(\tau) 
\label{plain2}
\end{eqnarray}
and the second is an $N$'th order polynomial equation in $t$ with coefficients 
which are polynomial in $x$ and $y$ and which also encode the dependence on 
the hypermultiplet mass $M$ and the $N-1$ moduli 
$u_{n}=\langle {\rm Tr}\Phi^{n}\rangle$, $n=2,\ldots,N$, of 
the four-dimensional Coulomb branch:
\begin{equation}
F_{N}(t,x,y: M, u_{n})=0
\label{simple2}
\end{equation}
$F_{N}$ also depends on $N-1$ additional parameters that we can think 
of as the coordinates on the fibres of a torus bundle over the
four-dimensional Coulomb branch. As in the $N=2$ case the total space 
of this fibration is a hyper-K\"{a}hler manifold ${\cal M}$ 
of complex dimension $2(N-1)$ which we identify as the Coulomb branch of the 
compactified theory \cite{KAP}.  
\paragraph{}  
Equation (\ref{simple2}) 
generalizes equation (\ref{simple}) of the $SU(2)$ case. 
Part of the problem here is that the $F_{N}$ 
grow rapidly in complexity with $N$ and there is no explicit formula 
in the general case. However let us review how much information we 
require to determine the exact superpotential. In the $N=2$ case we 
solved equations $(\ref{plain})$ and (\ref{simple}) eliminating 
$x$, $y$ in favour of $X$ and then $u$ in favour of $X$ and $P$. 
In particular, $X$ and $P$ provide an unconstrained parametrization 
of the target space. After solving the two constraint 
equations in this way, the corresponding F-terms vanish and 
only the mass term in the superpotential survives: ${\cal W}=\epsilon u(X,P)$. 
Hence the task of finding the superpotential is simply that of expressing 
the modulus $u=\langle {\rm Tr} \Phi^{2}\rangle$ in terms of a set of 
unconstrained holomorphic coordinates on ${\cal M}$. 
\paragraph{}
Fortunately, the information we need to accomplish this 
is provided by correspondence 
between ${\cal N}=2$ theories and integrable systems \cite{MM}. 
The particular version 
of this correspondence is the one given in \cite{KAP}, where the 
Coulomb branch of a compactified ${\cal N}=2$ theory 
is identified as the complexified phase space of a certain integrable 
Hamiltonian system. For a theory with gauge group $SU(N)$, 
the dynamical system consists of 
$N$ particles on the line interacting via 
two-body forces. The positions and momenta 
of the $N$ particles in the centre of mass 
frame are $X_{a}$ and $P_{a}$ $a=1,2,\ldots,N$. After implimenting the 
trivial centre of mass contraint 
$\sum_{a=1}^{N}X^{a}=\sum_{a=1}^{N}P^{a}=0$, the $N-1$ independent 
positions and momenta 
provide the unconstrained holomorphic parametrization of the Coulomb 
branch which we seek. 
\paragraph{}
The specific integrable 
system relevant for the ${\cal N}=2$ theory with an adjoint hypermultiplet 
considered above, is the elliptic 
Calogero-Moser Hamiltonian \cite{MA} 
defined by the two-body interaction potential 
$V(X)=M^{2}{\cal P}(X)$. The only fact we need here is that the $N-1$ Coulomb 
branch moduli $u_{i}$ are identified with the $N-1$ constants of motion of the 
system (where we are not counting the total linear momentum which we 
set to zero above). 
In particular the adjoint mass operator $u_{2}$ is identified 
with the first non-trivial constant of motion which is the 
Calogero-Moser Hamiltonian, ${\cal H}$ itself\footnote{
The quickest way to obtain this identification is from the equality 
demonstrated in \cite{DP} between ${\cal H}$ and 
$\partial {\cal F}/\partial\tau$, where 
${\cal F}(a_{n},m_{i},\tau)$ is the ${\cal N}=2$ prepotential 
written in terms of the `electric' periods $a_{n}$. We then invoke the 
Matone relation \cite{MAT,DKMM,DKM4,KMS} which, for UV finite theories, reads 
$u_{2}=\partial{\cal F}/\partial \tau$.}, 
\begin{equation}
u_{2}={\cal H}=\sum_{a=1}^{N} \frac{P_{a}^{2}}{2} + M^{2} \sum_{a>b} 
{\cal P}(X_{a}-X_{b})         
\label{cmh}
\end{equation}
Integrating out the momenta $P_{a}$, the superpotential becomes,  
\begin{equation}
{\cal W}= \epsilon M^{2} \sum_{a>b} {\cal P}(X_{a}-X_{b})         
\label{resn}
\end{equation}
which agrees with the proposal given in Section 3. 

\section{A Test for Gauge Group $SU(N)$}
\paragraph{}
In this Section we will perform a simple test of the proposed superpotential 
for the $SU(N)$ theory. 
In particular, we will compare the critical points 
of the superpotential with the vacua of the corresponding 
four dimensional theory. In the present case, 
the vacuum structure of the ${\cal N}=1$ theory 
which is obtained by soft breaking of ${\cal N}=2$ supersymmetry 
has been discussed in detail by Donagi and Witten \cite{DW}. 
As above the four-dimensional Coulomb branch is described by the two complex 
equations (\ref{plain2}) and (\ref{simple2}). The first equation describes 
the bare spectral curve which is a torus $E$ with complex 
structure parameter $\tau$. As in Section 3, this can be represented 
as the quotient 
$C/\Gamma$ where $\Gamma$ is a lattice in the complex plane: 
$\Gamma=Z\oplus \tau Z$. 
As the second equation is an $N$'th order polynomial equation in $t$ with 
coefficients which depend on $x$ and $y$ it defines an $N$-fold cover, 
$C\rightarrow E$. The four-dimensional Coulomb branch is then (part of) the 
Jacobian manifold of $C$. Away from the singular points in its moduli space, 
$C$ corresponds to a Riemann surface of genus $N$. 
\paragraph{}
As usual for ${\cal N}=2$ theories in four dimensions, 
the points on the Coulomb branch which survive as massive vacua after 
breaking to ${\cal N}=1$, correspond to points in moduli space where $C$ 
develops nodes and reduces to a surface of lower genus \cite{SW1}. 
In this case, 
the relevant singular points are those at which the maximal degeneration 
occurs and the genus of $C$ is reduced to one. 
According to Donagi and Witten, these points are in one to one correspondence 
with sublattices $\tilde{\Gamma}\subset \Gamma$ of index $N$. 
For the present purposes the index just means the number of points of 
$\Gamma$ contained in each period parallelogram of $\tilde{\Gamma}$. 
A particularly attractive aspect of this correspondence, is that the 
action of S-duality on the set, ${\cal S}$, of massive vacua is 
manifest. Specifically, the S-duality group is identified with  the 
$SL(2,Z)$ automorphism group of $\Gamma$. In general, $SL(2,Z)$ tranformations 
on ${\Gamma}$ will preserve the index of a sublattice $\Gamma'$ 
but not the sublattice itself. It follows that S-duality permutes 
the massive vacua. This is very natural, as each massive vacuum of the theory 
 is associated with the condensation of a particular BPS state of the 
 ${\cal N}=2$ theory. An action of $SL(2,Z)$ on the set of massive vacua
 is then inherited from the action of S-duality on the BPS spectrum.   
\paragraph{}
It is useful to note for each sublattice $\tilde{\Gamma}$ of index 
$N$ in $\Gamma$, the lattice $\tilde{\Gamma}/N$ itself contains $\Gamma$ as a 
sublattice of index $N$. In fact  $\tilde{\Gamma} \subset \Gamma$
with index $N$ if and only if $\Gamma \subset \tilde{\Gamma}/N$ 
with index $N$. If we now define $\Gamma'=\tilde{\Gamma}/N$ 
we can give an equivalent formulation of the correspondence.  
Namely, given the lattice 
$\Gamma$, the massive vacua are in one to one 
correspondence with lattices $\Gamma'$ such that $\Gamma$ is a sublattice 
of $\Gamma'$ with index $N$. 
\paragraph{}
In the previous section we discovered that the bare spectral curve 
$E=C/\Gamma$ is naturally identified with the torus on which 
the Weierstrass function is defined in the superpotential (\ref{resn}).   
In the following, we will demonstrate that ${\cal W}$ has a 
stationary point corresponding to each $\Gamma'$ which contains $\Gamma$ 
as a sublattice of index $N$. An intuitive way of understanding this result 
is suggested by the correspondence with the elliptic Calogero-Moser system 
described above. In particular, we can think of the complex scalars $X^{a}$ 
as the locations of $N$ particles on the torus $E$. 
The particles interact via a two-body `potential'  $\sim {\cal P}(X)$, 
albeit a complex one. The condition for a supersymmetric vacuum is 
just a kind of holomorphic version of the equilibrium condition for 
this $N$-body system. As the particles interact 
via complex repulsive `forces'\footnote{Of course the analogy is a 
loose one because the `force' between two particles is not directed 
along the line in the complex plane joining their positions.} 
$\sim {\cal P}'(X)$, the natural equilibrium 
positions are when the particles form a uniform lattice on $E$ 
so that the `forces' cancel pairwise. As there are $N$ particles, 
this defines a lattice $\Gamma'$ which has $\Gamma$ as a sublattice 
of index $N$.    
\paragraph{}
We will begin by considering the Weierstrass function, ${\cal P}(X:\Gamma)$,  
defined by the lattice $\Gamma=\{z=2m\omega_{1}+2n\omega_{2}: 
 (m,n)\in Z^{2} \}$. Two points which we will need are: 
\paragraph{}
{\bf 1:} The derivative of the Weierstrass function can be represented as, 
\begin{eqnarray}
{\cal P}'(X:\Gamma) & = & -2\sum_{Z\in \Gamma} \frac{1}{(X-Z)^{3}} 
\nonumber \\ & =& -2\sum_{(m,n)\in Z^{2}} \frac{1}{(X-2m\omega_{1}-2n\omega_{2})^{3}}
\label{pprime}
\end{eqnarray}
The resulting double 
series is uniformly convergent for all $X\notin \Gamma$. In particular, this 
means that the sum does not depend on which order the individual 
sums over $m$ and $n$ are performed \cite{EF}.   
 \paragraph{}
 {\bf 2:} For any lattice $\Gamma$, we know that a point $z=w$ is in $\Gamma$ 
 if and only the point $z=-w$ is. As the series in question converges, 
 we then have 
\begin{equation}
\sum_{Z\in \Gamma-\{0\}} \frac{1}{Z^{3}}=0
 \label{E3}
 \end{equation}
 \paragraph{}
 To demonstrate the required stationary points of ${\cal W}$, 
 must solve the $N$ equations\footnote{In this section we set all masses to 
 unity.}    
 \begin{equation}
 \frac{\partial {\cal W} }{\partial X_{b}} = 
 -\sum_{a\neq b} {\cal P}'(X_{a}-X_{b})=0
 \label{deriv}
 \end{equation}
 for $b=1,\ldots,N$ modulo lattice translations and the action of the 
 Weyl group. Using (\ref{pprime}), we have, 
 \begin{eqnarray} 
  \frac{\partial {\cal W}}{\partial X_{b}} & = & 2\sum_{a\neq b} \sum
  _{Z\in \Gamma} \frac{1}{(X_{a}-X_{b}-Z)^{3}} \nonumber \\
  & = & 2\sum_{a=1}^{N} \sum
  _{Z\in \Gamma-\{ 0\}} \frac{1}{(X_{a}-X_{b}-Z)^{3}}
 + 2 \sum_{a\neq b} \frac{1}{(X_{a}-X_{b})^{3}} \nonumber \\
 & =& \sum_{Y\in S(b)} \frac{1}{Y^{3}} +  \sum_{Y\in T(b)} \frac{1}{Y^{3}}
  \label{big1}
 \end{eqnarray}
 with the sets $S(b)$ and $T(b)$ defined as, 
  \begin{eqnarray}
 S(b) &= &\{Y=X_{a}-X_{b}-Z: Z\in \Gamma-\{0\}: a=1,2\ldots,N \} 
 \nonumber  \\ T(b) &= &\{Y=X_{a}-X_{b}: a=1,2\ldots,\hat{b},\ldots,N \} \\  
 \label{sets3}
 \end{eqnarray} 
 for each $b=1,2,\ldots , N$. The notation $\hat{b}$ appearing in the 
 definition $T(b)$ indicates that the $a=b$ term is omitted. As 
 each of the double sums considered is absolutely convergent, it suffices 
 to specify the set of points in $C$ over which each sum 
 is to be performed. In particular, we do 
 not need to specify the order in which the component summations 
 implicit in each term of (\ref{big1}) are to be taken.  
 Note also that, in the second line of 
 (\ref{big1}), we have used the fact that the $a=b$ term 
 in the first series on the 
 RHS vanishes because of (\ref{E3}).      
 \paragraph{}
 As above we will consider a second lattice $\Gamma'$ 
 such that $\Gamma \subset \Gamma'$ with index $N$. 
 By definition this means that there are $N$ points of $\Gamma'$ 
in each period 
 parallelogram $D_{\Gamma}$ of $\Gamma$: 
 $|\Gamma'\cap  D_{\Gamma}(z_{0})|=N$ for 
 all $z_{0}\in C$. Now we will consider a configuration such that  
 \begin{equation}
 \{X_{a}:a=1,2,\ldots N\}=\Gamma'\cap D_{\Gamma}(0) 
 \label{set2}
  \end{equation}
 Taking into account the action of the Weyl group, this choice corresponds
 to a single point on the Coulomb branch.  
\paragraph{}
 The idea now is that, for this configuration, 
 summing over $a$ and over points in $\Gamma$ is equivalent to summing 
 over points in $\Gamma'$. To see this note that we have, 
 \begin{equation} 
 \Gamma'=\{Y=X_{a}-Z: Z\in \Gamma, a=1,2,\ldots,N\}
 \label{gprime}
 \end{equation}
 The lattice is invariant under translations by any lattice vector. Hence,  
 as $X_{b}\in \Gamma'$, we also have, 
 \begin{equation} 
 \Gamma'=\{Y=X_{a}-X_{b}-Z: Z\in \Gamma, a=1,2,\ldots,N\}
 \label{gprime2}
 \end{equation}   
 for each $b$. From the definitions of $S(b)$ and $T(b)$ we now see that 
  \begin{equation} 
  S(b)\cup T(b)=\Gamma' -\{0\}
  \label{sus}
  \end{equation}
  For each $b=1,2,\dots N$. Finally, using this relation and 
   (\ref{E3}), we have 
  \begin{eqnarray} 
  \frac{\partial {\cal W}}{\partial X_{b}} & = & 
  \sum_{Y\in S_{1}} \frac{1}{Y^{3}} +  
 \sum_{Y\in S_{2}} \frac{1}{Y^{3}} \nonumber \\
 & = &  \sum_{Y\in \Gamma'-\{0\}} \frac{1}{Y^{3}} =0
 \label{big2}
 \end{eqnarray}
  This shows that ${\cal W}$ has a critical point 
  corresponding to each lattice $\Gamma'$ which has $\Gamma$ 
  as a sublattice of index $N$. Hence the superpotential reproduces the 
  set of vacua predicted by the Donagi-Witten curve. There are several  
  loose ends here: we have not shown that the critical points found above are 
  non-degenerate and thus correspond to massive vacua.  
   Nor have we shown that they are unique. Finally, apart for the explicit 
  calculation for gauge group $SU(3)$ reported in Section 3, we have not 
  checked the existence of the massless vacua predicted by the 
  classical analysis of Section 2. These issues are under investigation.  
  \paragraph{}
    In the next Section we will need to know the explicit form for the vacuum 
    configurations in the case where $N$ is prime.  
     In this case, by the analysis of Section 1.1, we should find $N$ vacua 
   in the confining phase and a single vacuum in the Higgs phase. 
   As above these each of these vacua should correspond to 
   lattices $\Gamma'$ such that $\Gamma\subset \Gamma'$ with index $N$. 
   If $\Gamma$ has periods $(2\omega_{1},2\omega_{2})$ then two obvious 
   solutions of this condition are the lattices $\Gamma_{A}'$ with 
   periods $(2\omega_{1}/N,2\omega_{2})$ and   $\Gamma_{B}'$ with 
   periods $(2\omega_{1},2\omega_{2}/N)$.    
   From Section 1.1, 
   the torus $E=C/\Gamma$ has periods $2\omega_{1}=2\pi i$ and 
   $2\omega_{2}=2\pi i \tau$. 
   Thus, the two vacuum configurations corresponding to 
   $\Gamma'_{A}$ and $\Gamma'_{B}$ are,  $X_{a}^{(A)}=2\pi ia /N$  and 
   $X^{(B)}_{a}=2\pi i \tau a/N$ respectively, 
   with $a=1,2\ldots N$ in both cases. 
   The coupling independence of configuration A identifies 
   it as the vacuum of Higgs sector: this is the only vacuum which is 
   visible in the semiclassical limit $\tau\rightarrow\infty$. The 
   second vacuum must therefore be one of the $N$ confining vacua. 
   From the point of view of the four-dimensional ${\cal N}=2$ theory, vacua
    $A$ and $B$ are associated with the condensation of elementary quanta and 
    magnetic monopoles respectively. The 
   remaining $N-1$ confining vacua are generated by the modular 
   transformation $\tau\rightarrow \tau +p$ for $p=1,2,\ldots,N$. 
   In the ${\cal N}=4$ theory, this transformation turns a magnetic monopole 
   into a dyon which carries electric charge $p$. Correspondingly, for 
   soft breaking of ${\cal N}=2$ SUSY these 
   vacua are associated with the condensation of massless BPS dyons. 
  
  \section{An Application}
  \paragraph{}
  A simple application of the superpotential derived above is to calculate 
  the condensates of gauge invariant chiral operators. 
  An obvious candidate is the value of the modulus  
  $u_{2}=\langle {\rm Tr} \Phi^{2} \rangle$ 
  in each vacuum. According to the results of Section 4, we have 
  $u_{2}={\cal W}/\epsilon$, thus the vacuum values of $u_{2}$ are given 
  by the critical values of the superpotential. There is a subtlety with this 
  identification which is best illustrated in the $SU(2)$ case. Recall that  
  the three critical values of the superpotential were found in Section 
  3 to be $e_{1}(\tau)$, $e_{2}(\tau)$ and $e_{3}(\tau)$. These three 
  functions of $\tau$ transform under the modular group with weight two 
  modulo permutations. However this result depended on setting the additive 
  constant $C(\tau)$ appearing in (\ref{sp1}) to zero. In particular we 
  noted that any non-zero value for $C(\tau)$ would spoil the modular property 
  of the superpotential. In fact this precisely matches a similar ambiguity 
  in the parameter $u$ discussed in \cite{DKM4}. 
  If we retain the notation 
  $u$ for the classical modulus $\langle {\rm Tr} \Phi^{2} \rangle$, 
  the Seiberg-Witten curve can be parameterized by a coordinate 
  $\tilde{u}$ which transforms with modular weight two. On general grounds 
   the two coordinates are related as\footnote{I would like to thank Matt 
   Strassler for pointing out the incorrect discussion of this point given 
   in a previous version of this paper.}, 
  \begin{equation} 
   \tilde{u}=u-M^{2}\left
   (\frac{1}{12}+\sum_{p=1}^{\infty}\alpha_{p}q^{p}\right)
   \label{utilde}   
   \end{equation}
   where the coefficients $\alpha_{p}$ are unknown for $p>1$. 
   If we choose $C(\tau)=0$, the corresponding superpotential of modular 
   weight two is therefore ${\cal W}=\epsilon \tilde{u}$. In view of 
 (\ref{utilde}), predictions 
   for $\langle {\rm Tr} \Phi^{2} \rangle$ in different vacua are only 
   determined up to an overall additive function of $\tau$. Thus, only 
   predictions for differences between the values
   of $u$ in distinct vacua are unamibguous.       
   A similar 
   additive ambiguity arises in the definition of $u_{2}$ 
   for $N>2$ but we will suppress it in the following.  
   Note that the critical values of 
  ${\cal W}$ also determine the gluino condensate via 
  $\langle {\rm Tr}\lambda_{\alpha}\lambda^{\alpha} \rangle =
  \partial_{\tau}{\cal W}$.  As the superpotential is independent of $R$,
   the results will apply equally in the four-dimensional limit. 
   \paragraph{}
   In the following we will restrict our attention to the 
   massive vacua and then to the 
   simplest case of prime $N$. To compute the critical value of the 
   superpotential in the Higgs vacuum given in the previous section, we 
   must compute the sum, 
   \begin{eqnarray}
   u^{(A)}_{2}/M^{2}=\left. {\cal W} \right|_{X=X^{(A)}}  & = 
    & \frac{1}{2}\sum_{b=1}^{N} 
   \sum_{a \neq b}^{N} {\cal P}\left( \frac{2(a-b)}{N}i \pi\right)  
   \nonumber \\ & = & \frac{N}{2}\sum_{d=1}^{N-1} {\cal P}\left(
   \frac{2\pi id}{N}\right)  
   \label{sum}
   \end{eqnarray}     
   As we have $\omega_{1}=i\pi$ and $\omega_{2}=i\pi \tau$, the Weierstrass 
   function can be written as,   
   \begin{equation}
   {\cal P}(2\pi i z)=-\frac{1}{4\pi^{2}}\left( \frac{1}{z^{2}} 
   + \sum _{(m.n)\neq (0,0)}\left[\frac{1}{(z-m-n\tau)^{2}}-
   \frac{1}{(m+n\tau)^{2}} \right] \right)
   \label{weier2}
   \end{equation}
   The double series on the RHS is absolutely convergent and therefore 
   independent of the order of summation over $m$ and $n$. However, some care 
   is required because, when taken in isolation, either of the two terms 
   in the summand leads to a double series which is only conditionally 
   convergent and, in particular, depends the order of summation. To 
   illustrate this we consider the second term which, for one choice 
   of ordering, is proportional the regulated second Eisenstein 
   series \cite{KOB}, 
   \begin{equation}
   E_{2}(\tau)=\frac{3}{\pi^{2}}
   \sum_{n=-\infty}^{+\infty} {\sum_{m=-\infty}^{+\infty}}'
      \frac{1}{(m+n\tau)^{2}}
    \label{eis}
    \end{equation}
    The prime on the $m$ summation means `omit the $m=0$ term when $n=0$'.
    If we naively interchange the order of summation and 
    multiply both sides by $\tau^{2}$ we would obtain the relation 
    $E_{2}(-1/\tau)=\tau^{2}E_{2}(\tau)$. As $E_{2}$ is also invariant under 
    $\tau\rightarrow \tau + 1$, this would lead us to conclude 
    incorrectly that $E_{2}$ is a modular form of weight two.  
    In fact the summations over $m$ and $n$ fail
    to commute and instead give an anomalous transformation 
    law for $E_{2}$ under the modular transformation 
    $\tau\rightarrow -1/\tau$, 
    \begin{equation}
    E_{2}(\tau)=\frac{1}{\tau^{2}}E_{2}\left(-\frac{1}{\tau}\right)-
    \frac{6}{\pi i\tau}
    \label{anomaly}
    \end{equation}
     \paragraph{}
    Paying attention to the order of summation we obtain, 
   \begin{eqnarray}
    u_{2}^{(A)}/M^{2}  & = &  
   -\frac{N}{8\pi^{2}}\left[\sum_{d=1}^{N-1} \frac{N^{2}}{d^{2}}
   + \sum_{n=-\infty}^{+\infty} {\sum_{m=-\infty}^{+\infty}}'
   \left(\left[\sum_{d=1}^{N} \frac{N^{2}}{(mN-d+nN\tau)^{2}}\right]-
   \frac{N}{(m+n\tau)^{2}}\right)\right] \nonumber \\
   &=& -\frac{N}{8\pi^{2}}\left[
   \hat{\sum_{l=-\infty}^{+\infty}} \frac{N^{2}}{l^{2}}
   +  \hat{\sum_{n=-\infty}^{+\infty}} \sum_{l=-\infty}^{+\infty}
    \frac{N^{2}}{(l+nN\tau)^{2}}-\frac{N\pi^{2}}{3}E_{2}(\tau) \right] 
    \nonumber \\
    & = & \frac{N^{2}}{24} \left(E_{2}(\tau)-NE_{2}(N\tau)\right) 
  \label{sum4}
   \end{eqnarray}
   where the term where the summation variable equals zero is omitted in 
   each of the hatted sums.   
   A similar calculation yields the expectation value of $u_{2}$ in the 
   confining vacuum corresponding to the lattice $\Gamma'_{B}$ as, 
   \begin{eqnarray}
   u^{(B)}_{2}/M^{2} &= & \left. {\cal W} \right|_{X=X^{(B)}}  \nonumber \\
        & = & \frac{N^{2}}{24}\left(E_{2}(\tau)-
        \frac{1}{N}E_{2}\left(\frac{\tau}{N}\right)\right) 
  \label{uconf} 
  \end{eqnarray}
  The value of $u_{2}$ in the remaining 
  $N-1$ confining vacua is obtained by replacing $\tau$ with 
  $\tau+p$ in (\ref{uconf}) for $p=1,2,\ldots, N-1$. 
  Using equation (\ref{anomaly}) we find that $u_{2}^{(A)}(-1/\tau)=\tau^{2}
  u_{2}^{(B)}(\tau)$. Notice that the anomalous term in the transformation 
  law for $E_{2}$ cancels between the two terms contributing to $u_{2}^{(A)}$. 
  Hence we find that the vacuum values of $u_{2}$ tranform with weight two 
  under, and are also interchanged by, an electric-magnetic duality 
  transformation. This generalizes the $SU(2)$ result for $\tilde{u}$ 
  described above. The 
  gluino condensate $S=\langle {\rm Tr} \lambda_{\alpha}\lambda^{\alpha}
  \rangle$ in each vacuum is obtained by a differentiating the 
  corresponding expression for $u_{2}$ with respect to $\tau$.  
  As this differentiation does not commute with a general $SL(2,Z)$ 
  tranformation, the resulting functions of $\tau$ do not have any 
  special modular properties. Finally, we note that this calculation 
  can easily be generalized to the condensates $u_{n}$ with $n>2$ 
  \cite{WIP}.      
  \section{Conclusion}
  \paragraph{}
  The main result of this paper is that the exact superpotential of 
  the ${\cal N}=1$ theory obtained by introducing chiral multiplet masses 
  in ${\cal N}=4$ supersymmetric $SU(N)$ Yang-Mills theory on 
  $R^{3}\times S^{1}$ is,  
  \begin{equation} 
  {\cal W}= m_{1}m_{2}m_{3}\, \sum_{a>b} \, {\cal P}(X_{a}-X_{b})
  \label{sp11}
  \end{equation} 
  An interesting feature of this result is that the superpotential coincides 
  with the complexified potential of the elliptic Calogero-Moser system. 
  The same integrable system is associated 
  with ${\cal N}=2$ supersymmetric Yang-Mills theory with one adjoint 
  hypermultiplet of mass $M$. 
  Specifically the spectral curve which encodes the 
  conserved quantities of the Calogero-Moser system is the same 
  complex curve which governs the Coulomb branch of the four-dimensional  
  ${\cal N}=2$ theory. This suggests a new connection between 
  ${\cal N}=1$ theories and integrable systems which extends the 
  correspondence between ${\cal N}=2$ theories and integrable systems 
  introduced in \cite{DW,MM}. Starting from an ${\cal N}=2$ 
  theory in four dimensions, the superpotential of the ${\cal N}=1$ 
  theory on $R^{3}\times S^{1}$ obtained by introducing a mass for the 
  adjoint scalar in the ${\cal N}=2$ vector multiplet should coincide   
  with the complexified potential of the corresponding integrable system.  
  This connection should 
  be a general one which applies to all ${\cal N}=1$ theories obtained by 
  soft breaking of ${\cal N}=2$ supersymmetry. 
  Following the arguments given in Section 4,   
  the correspondence can be `explained' by combining two previous 
  observations: 
\paragraph{}
{\bf 1:} The Coulomb branch of an ${\cal N}=2$ theory on $R^{3}\times S^{1}$ 
coincides with the complexified phase space of the integrable Hamiltonian 
system associated with the corresponding theory on $R^{4}$ \cite{KAP}. 
The dynamical variables of the integrable system, $\{X^{a},P^{a}\}$ 
yield holomorphic 
coordinates on the Coulomb branch of the compactified theory.  
\paragraph{}
{\bf 2:} The complex moduli $u_{n}$, $n=2\ldots N$, which 
 parametrize the four dimensional Coulomb branch correspond to the 
 conserved quantities of the integrable system \cite{DW}. In particular, 
 we may identify $u_{2}$ with the Hamiltonian, ${\cal H}(X^{a},P^{a})$, 
 of the integrable system (see Footnote 3 above). 
 Introducing a superspace mass term  
${\cal W}=\epsilon u_{2}$ for the adjoint scalar in the ${\cal N}=2$ 
vector multiplet, we obtain an ${\cal N}=1$ theory on $R^{3}\times S^{1}$ 
with superpotential ${\cal W}=\epsilon{\cal H}(X^{a},P^{a})$. As the momenta 
appear via the non-relativistic kinetic terms
\footnote{This point would need to be modified if the correspondence is 
to apply also to ${\cal N}=2$ theories obtained by compactification on 
a circle from five dimensions \cite{NEK}, where 
the corresponding integrable systems are relativistic.}, 
$\sum P_{a}^{2}/2$, they can 
be trivially eliminated via their equations of motion and the resulting 
superpotential is just the complexified potential of the 
integrable system.   
\paragraph{}  
One case where we can immediately make contact with known results is 
for the ${\cal N}=2$ theory without adjoint hypermultiplets. For gauge 
group $SU(N)$, the relevant integrable system is the affine Toda lattice 
\cite{MM}. 
The exact superpotential of the corresponding ${\cal N}=1$ theory on 
$R^{3}\times S^{1}$ is known to coincide with the affine Toda potential 
\cite{KV}. In fact, this superpotential can also be obtained directly from 
(\ref{sp11}) by decoupling the three adjoint chiral multiplets. 
The appropriate limit, which is well known on the integrable system side of 
the correspondence \cite{IN}, generalizes the limit discussed in Section 3 
for the $SU(2)$ case.
It would be interesting to test this correspondence for other 
${\cal N}=2$ theories for which the corresponding integrable system 
is known. The correspondence may also suggest new results in cases where  
the relevant integrable system is unknown. 
\paragraph{}
I would like to thank David E. Kaplan and Prem Kumar for useful discussions 
as well as Kimyeong Lee and Dave Tong for reading a preliminary draft of 
this paper. 
I am gratefull to the particle theory group at SISSA for hospitality 
during the completion of this work. The author also acknowledges the support 
of TMR network grant FMRX-CT96-0012.

\end{document}